%% file: sample-manuscript.tex
\documentclass[acmsmall, screen]{acmart}

\AtBeginDocument{%
  \providecommand\BibTeX{{%
    \normalfont B\kern-0.5em{\scshape i\kern-0.25em b}\kern-0.8em\TeX}}}

\setcopyright{acmcopyright}
\copyrightyear{2023}
\acmYear{2023}
\acmDOI{XXXXXXX.XXXXXXX}

\usepackage{multirow}
\usepackage[skins]{tcolorbox}

\usepackage{subcaption}
\usepackage{xspace}

\newcommand{\tool}{\textsc{Mar\-Mot}\@\xspace} 

\usepackage{tikz}
\usepackage{xcolor}
\newcommand*\circled[1]{\tikz[baseline=(char.base)]{
            \node[shape=circle,fill,inner sep=0.5pt] (char) {\textcolor{white}{#1}};}}

\usepackage{float}
\newtcolorbox{custombox}[1]{
	colback=gray!10,
	colframe=gray!20,
	left=1mm,
	right=1mm,
	top=1mm,
	bottom=1mm,
	fonttitle=\bfseries,
	arc=2mm,
	leftrule=0mm,
	rightrule=.5mm,
	toprule=0mm,
	bottomrule=.5mm,
	notitle,
	before=\par\smallskip\noindent,
	before upper={\textbf{#1: } },
}
\usepackage{colortbl}
\newsavebox\CBox
\def\textBF#1{\sbox\CBox{#1}\resizebox{\wd\CBox}{\ht\CBox}{\textbf{#1}}}
\usepackage{chngcntr}

\begin{document}

\title{\tool: Metamorphic Runtime Monitoring of Autonomous Driving Systems}

\author{Jon Ayerdi}
\affiliation{%
  \institution{Mondragon University}
  \streetaddress{Goiru 2}
  \city{Mondragon}
  \state{Gipuzkoa}
  \country{Spain}
  \postcode{20500}
}
\email{jayerdi@mondragon.edu}

\author{Asier Iriarte}
\affiliation{%
  \institution{Mondragon University}
  \streetaddress{Goiru 2}
  \city{Mondragon}
  \state{Gipuzkoa}
  \country{Spain}
  \postcode{20500}
}
\email{airiartem@mondragon.edu}

\author{Pablo Valle}
\affiliation{%
  \institution{Mondragon University}
  \streetaddress{Goiru 2}
  \city{Mondragon}
  \state{Gipuzkoa}
  \country{Spain}
  \postcode{20500}
}
\email{pvalle@mondragon.edu}

\author{Ibai Roman}
\affiliation{%
  \institution{Mondragon University}
  \streetaddress{Goiru 2}
  \city{Mondragon}
  \state{Gipuzkoa}
  \country{Spain}
  \postcode{20500}
}
\email{iroman@mondragon.edu}

\author{Miren Illarramendi}
\affiliation{%
  \institution{Mondragon University}
  \streetaddress{Goiru 2}
  \city{Mondragon}
  \state{Gipuzkoa}
  \country{Spain}
  \postcode{20500}
}
\email{millarramendi@mondragon.edu}

\author{Aitor Arrieta}
\affiliation{%
  \institution{Mondragon University}
  \streetaddress{Goiru 2}
  \city{Mondragon}
  \state{Gipuzkoa}
  \country{Spain}
  \postcode{20500}
}
\email{aarrieta@mondragon.edu}

\renewcommand{\shortauthors}{Ayerdi et al.}

\begin{abstract}
\input{Abstract}
\end{abstract}



\maketitle

\section{Introduction}
\input{Introduction}

\section{Background}
\label{sec:Background}
\input{Background}

\section{Approach}
\label{sec:approach}
\input{Approach}

\section{Evaluation Setup}
\label{sec:Evaluation}
\input{Evaluation}

\section{Related Work}
\label{sec:relatedwork}
\input{RelatedWork}

\section{Conclusion}
\label{sec:conclusion}
\input{Conclusion}

\begin{acks}
This work was partially founded by the Basque Government through their Elkartek program (EGIA project, ref. KK-2022/00119 and SIIRSE project, ref. KK-2022/00007). The authors are part of the Software and Systems Engineering research group of Mondragon Unibertsitatea (IT1519-22), supported by the Department of Education, Universities and Research of the Basque Country.

\end{acks}

\bibliographystyle{ACM-Reference-Format}
\bibliography{sample-base}

\appendix
\counterwithin{figure}{section}
\section{Uncertainty distributions under nominal conditions}
\label{apx:uncertaintydist}
\input{appendix1}

\end{document}

%% file: Abstract.tex
Autonomous Driving Systems (ADSs) are complex Cyber-Physical Systems (CPSs) that must ensure safety even in uncertain conditions. Modern ADSs often employ Deep Neural Networks (DNNs), which may not produce correct results in every possible driving scenario. Thus, an approach to estimate the confidence of an ADS at runtime is necessary to prevent potentially dangerous situations. In this paper we propose \tool, an online monitoring approach for ADSs based on Metamorphic Relations (MRs), which are properties of a system that hold among multiple inputs and the corresponding outputs. Using domain-specific MRs, \tool estimates the uncertainty of the ADS at runtime, allowing the identification of anomalous situations that are likely to cause a faulty behavior of the ADS, such as driving off the road.

We perform an empirical assessment of \tool with five different MRs, using two different subject ADSs, including a small-scale physical ADS and a simulated ADS. Our evaluation encompasses the identification of both external anomalies, e.g., fog, as well as internal anomalies, e.g., faulty DNNs due to mislabeled training data. Our results show that \tool can identify up to 65\% of the external anomalies and 100\% of the internal anomalies in the physical ADS, and up to 54\% of the external anomalies and 88\% of the internal anomalies in the simulated ADS. With these results, \tool outperforms or is comparable to other state-of-the-art approaches, including SelfOracle, Ensemble, and MC Dropout-based ADS monitors.


%% file: Introduction.tex
In the last few years, the autonomy of vehicles is increasing, in part, thanks to the recent advances of machine-learning based technologies. Deep Neural Networks (DNNs) have enabled Autonomous Driving Systems (ADSs) to operate in dynamic environments with little to no human intervention~\cite{bojarski2016end}. Supervised training is a particularly effective approach that allows DNNs to effectively learn driving behaviors (such as the throttling or the steering angle) from labeled datasets of camera images and the corresponding expected driving behavior~\cite{bojarski2016end}. However, to deploy ADSs in the roads, these should be reliable, even in the presence of uncertain or unforeseen scenarios. To increase the reliability of these systems, in the last few years, significant effort has been devoted to research on automated test case generation techniques that falsify safety properties of ADSs (e.g., find collisions, violate traffic laws)~\cite{tian2018deeptest,haq2022efficient,haq2023many,abdessalem2018testing,abdessalem2018testing2,ben2016testing,zhong2022neural,lu2022learning,zhou2023specification,sun2022lawbreaker,scheuer2023stretch,calo2020generating,gambi2019automatically, gambi2019generating,biagiola2022testing}. However, the expensive execution of test cases (due to the use of simulation frameworks), as well as the huge input space, makes it impossible to cover all situations in which the ADS can misbehave.

To address this issue, existing works propose runtime monitors (also known as supervisors or failure predictors) to assess the level of the DNN dependability in operation~\cite{henriksson2019towards,hussain2022deepguard,kim2019guiding,stocco2020misbehaviour,stocco2022confidence,stocco2022thirdeye,wang2020dissector,xiao2021self,zhang2018deeproad}. In the context of ADSs, both, black-box as well as white-box approaches have been proposed. As for black-box approaches, techniques like SelfOracle~\cite{stocco2020misbehaviour} or DeepGuard~\cite{hussain2022deepguard} monitor the ADS by examining its behavior in response to the input images of the system. In the context of ADS, for black-box runtime monitoring techniques, one of the identified core limitations~\cite{stocco2022thirdeye} relates to the impossibility of handling other failures than those caused due to data-driven bugs, e.g., failures caused by corrupted images. Therefore, these techniques fail to detect faults due to inadequate training or by bugs at the DNN model level~\cite{stocco2022thirdeye}. To address this challenge, Stocco et al.~\cite{stocco2022thirdeye} proposed ThirdEye, a white-box ADS failure predictor that relies on attention maps produced by explainable artificial intelligence (XAI) techniques. Specifically, they propose three different summarization methods to compute a confidence score from the attention map of the images, provided some extra information previously extracted from the training dataset. Although this technique is promising and overcomes the limitations of black-box techniques, it also poses one core limitation that may prevent its adoption in practice: its computational cost, and as a direct consequence, its execution time. Specifically, the techniques employed to obtain the attention maps require several invocations of the DNN, which may make this technique unfeasible unless significant computational resources are available exclusively for the ADS monitors. This may be a problem in the context of ADS that use resource-constrained embedded devices.

To deal with the above-mentioned challenges, we propose \tool, a simple, yet effective and efficient runtime monitoring approach. \tool leverages ideas from metamorphic testing~\cite{segura2016survey} to predict when the ADS will fail, and it provides significant benefits over the state-of-the-art approaches. Firstly, based on our experiments with a lane keeping case study system, {\tool} provides \textbf{higher effectiveness} in terms of detecting internal (faulty DNN models) or external (unexpected inputs) anomalies which may cause an incorrect behavior of the system. Secondly, our approach is capable of an \textbf{earlier detection} of internal anomalies than other approaches, which is crucial in cases where there is only a short time to react. 
Lastly, our approach only requires cheap image transformations and one additional invocation of the ADS controller (either DNN or other), which would solve the high computational cost of some white-box techniques (e.g., ThirdEye~\cite{stocco2022thirdeye}).

Our paper makes the following contributions: 

\begin{itemize}
    \item \textbf{Technique: }We propose a black-box approach inspired on Metamorphic Testing to estimate the uncertainty of DNN-based ADSs at runtime. We define several image-based Metamorphic Relations (MRs) for this domain, and a process to evaluate them at runtime. Our technique can be used to identify potentially unexpected scenarios, either those derived from internal (e.g., improperly trained DNN) or external (e.g., fog blocking the visibility of the camera) uncertainties.

    \item \textbf{Evaluation: }Unlike previous studies, which use simulation environments in their evaluation, we employ a physical (small-scale) ADS. This was considered to reduce the validity threat related to the reality gap existing between simulated and real environments (e.g., due to the textures of the images or the fidelity of the simulation engines). With this physical vehicle, we carry out an empirical study, showing \tool's benefits over the black-box approach of SelfOracle~\cite{stocco2020misbehaviour} as well as the white-box Ensemble and MC Dropout approaches. We also complement our evaluation with datasets generated with simulation environments from previous studies \cite{stocco2022thirdeye}.

    \item \textbf{Dataset: }We provide the dataset of images used to train our DNN-based driving models, as well as the one used to evaluate our test oracles in this paper \cite{MarmotDataset}. Our dataset comprises images from a training track provided by the vendor of the mobile robot~\cite{LeoRoverDataset}, as well as images from another training track and two additional evaluation tracks provided by us. For the two evaluation tracks, our dataset includes recordings of correct driving under nominal conditions and internal (faulty DNN models) or external (unexpected inputs) anomalies, as well as DNN misbehaviors (out-of-bounds scenarios) under internal or external anomalies.

\end{itemize}

The rest of the paper is structured as follows: We provide basic background in Section \ref{sec:Background}. We explain our approach in Section \ref{sec:approach}. We describe our empirical evaluation setup in Section \ref{sec:Evaluation}, which includes a case study with a small-scale physical ADS, and a different simulation-based case study. We analyze the results from our experiments in Section \ref{sec:Evaluation:Results}. We present a general discussion with the key takeaways from our evaluation in Section \ref{sec:discussion}. We position our approach with the current state-of-the-art in Section \ref{sec:relatedwork}. Lastly, we conclude and discuss future research avenues in Section \ref{sec:conclusion}.

%% file: Background.tex
\subsection{Autonomous Driving Systems}

Modern ADSs leverage machine-learning technologies, such as DNNs, to analyze the inputs (e.g., camera images, LIDAR sensors, GPS) and drive the vehicle in real time with various degrees of autonomy~\cite{bojarski2016end}. Typically, DNNs learn tasks such as lane keeping in a supervised manner, i.e., by generalizing from a dataset of samples labeled by humans, which demonstrate correct driving behaviors~\cite{bojarski2016end}.

However, DNN-based driving systems are currently unlikely to generalize correctly for all possible driving conditions that can be encountered in a real road. Thus, in order to ensure safety, it is paramount to monitor the (estimated) uncertainty of the DNN, so that an appropriate healing strategy can be applied in high-uncertainty scenarios. Healing strategies to leave the vehicle in a safe state may include passing control to a human driver, reducing the speed, or completely stopping the vehicle~\cite{hussain2022deepguard}.

\subsection{Monitoring Techniques for ADSs}

An approach to monitor a DNN's behaviour and quantify its uncertainty level must consider several possible factors. On the one hand, this includes external factors, such as adverse environmental conditions like fog or heavy rain~\cite{tian2018deeptest,stocco2020misbehaviour,hussain2022deepguard}. On the other hand, internal factors such as poorly trained DNNs (e.g., wrong hyperparameter values, incorrect labels in the training dataset) may also need to be considered~\cite{humbatova2020taxonomy,stocco2022thirdeye}.

Black-box techniques estimate the uncertainty of the DNN based solely on its inputs (images) and outputs (steering angle, throttle, etc.), and possibly also the training data~\cite{stocco2020misbehaviour}. The main advantage of these techniques is that they are independent from the DNN architecture, and often even generalizable to systems which do not even employ DNNs. Nevertheless, their lack of knowledge of the DNNs internal behavior might limit the ability of these techniques to react to internal uncertainties~\cite{stocco2022thirdeye}. A popular black-box approach for estimating the uncertainty of DNNs in the context of ADSs is using autoencoders, which are DNNs that can reconstruct an input image, to identify inputs beyond the distribution of the inputs with which the ADS has been trained~\cite{stocco2020misbehaviour,hussain2022deepguard}. Metamorphic testing is an alternative black-box technique that can be used to identify inconsistencies in an ADS without interacting with its internals~\cite{tian2018deeptest,zhang2018deeproad}.

Conversely, white-box techniques monitor the internal behavior of the DNN, require a transparent access to the network, and are often specifically designed for networks with specific features. However, the knowledge of the DNN's internal behavior may allow these techniques to identify internal uncertainties that black-box techniques cannot. For example, with Deep Ensemble Neural Networks, multiple DNN models with different weight initialization or hyperparameters are trained with the same dataset to obtain a probability distribution of the model output~\cite{lakshminarayanan_simple_2017}. This distribution can then be used to infer the uncertainty of the DNN for the corresponding inputs~\cite{lakshminarayanan_simple_2017}.

\subsection{Metamorphic Testing}

Metamorphic Testing (MT) is an alternative technique to alleviate the test oracle problem that exploits known input and output relations that should hold among \emph{multiple} test executions, the so-called Metamorphic Relations (MRs)~\cite{chen1998metamorphic}. As an example, if the input image for an image-based classifier DNN 
is altered by slightly increasing its brightness, the output of the DNN should remain almost unchanged. Here, the MR we define consists of an \emph{Input Relation (IR)} ($img_2 = brighten(img_1)$) and an \emph{Output Relation (OR)} ($output_2 \simeq output_1$).

Typically, MT is performed with pairs of test cases: A \emph{source} test case, which is usually generated by a test generation strategy (e.g., random test generation), and a \emph{follow-up} test case, generated by \emph{transforming} the source test case in a way that the input relation of the MR is satisfied~\cite{segura2016survey,chen1998metamorphic}. For the given example, the follow-up image can be generated by brightening the source image. A violation of this MR (high deviation of the output between the source and follow-up images) might indicate that the DNN is not operating reliably, possibly due to uncertain conditions~\cite{weiss2023uncertainty}. Note that the satisfaction of a MR does not necessarily guarantee that the behavior of the system is correct.

This technique has already been applied in various forms for testing CPSs, such as autonomous drones~\cite{lindvall2017metamorphic} and driverless cars~\cite{zhou2019metamorphic}, including DNN-based autonomous driving systems \cite{tian2018deeptest,zhang2018deeproad}. Nevertheless, the existing research of this technique focuses on testing, as opposed to other verification tasks such as the online monitoring of systems.

%% file: Approach.tex
In this paper we propose \tool, a metamorphic runtime monitoring technique for ADS. We specifically focus on the use of this technique to provide a runtime monitor for DNNs-based lane-keeping ADSs. This runtime monitor will provide an uncertainty score for the system in real-time, which is expected to be correlated with the presence of anomalies that may cause system failures in the near future. Therefore, this approach can enable the real-time prediction of misbehaviors via MRs in order to take corrective actions before these occur.

\subsection{Architecture} \label{sec:Approach:Architecture}

\begin{figure*}[th!]
    \centering
    \includegraphics[width=.95\linewidth]{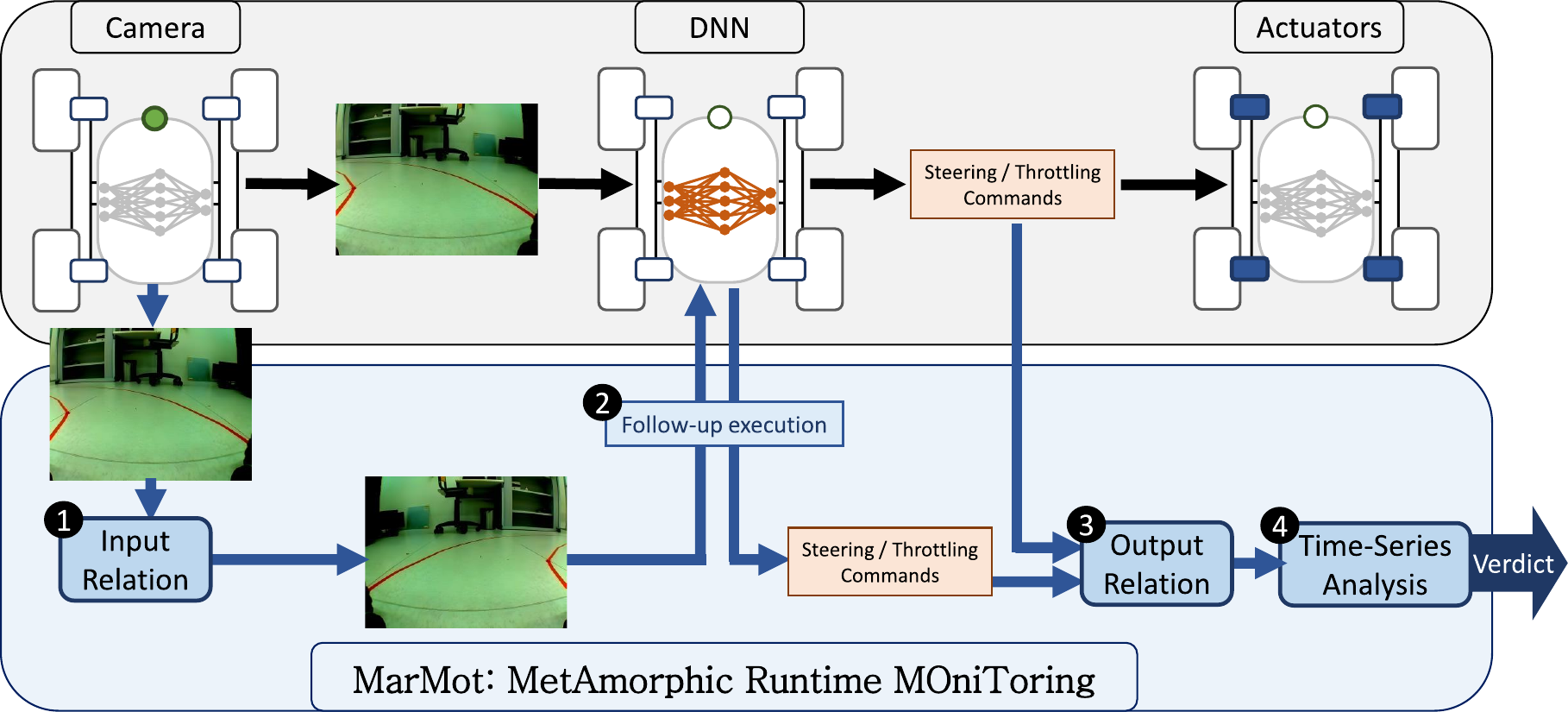}
    \caption{Architecture of \tool for DNN-based ADS }
    \label{fig:MetamorphicRuntimeLeoRover}
\end{figure*}

Figure~\ref{fig:MetamorphicRuntimeLeoRover} shows an overview of the architecture of \tool, which we use for applying Metamorphic Runtime Monitoring to an ADS controlled by an image-based DNN. The top half of the image shows the three main components of the ADS: (1) the camera, for the image acquisition; (2) the DNN, for decision-making based on the images; and (3) the actuator controls, which act on the steering and throttling commands provided by the DNN.

The bottom half of the image shows the Metamorphic Runtime Monitoring components and their interactions. For each MR used by our monitor, an Input Relation (IR) and its corresponding Output Relation (OR) must be defined.

Firstly, the Input Relation \circled{1} defines a transformation of the input image. In the example from Figure~\ref{fig:MetamorphicRuntimeLeoRover}, the IR consists of flipping the image horizontally, i.e., $img_f = flip(img_s)$. By applying this transformation to the image captured by the camera (i.e., $img_s$), which we consider the \emph{source} input, we generate the \emph{follow-up} input for the MR (i.e., $img_f$).

After the \emph{follow-up} image has been generated, it is executed by the DNN to obtain its output \circled{2}. This step corresponds with the execution of the follow-up test case in Metamorphic Testing, which is possible because the DNN is usually an isolated component of the ADS, with no direct dependencies on external states. It is important to note that the output of the DNN for the follow-up image is not sent to the driving commands, but instead used for monitoring purposes.

Finally, the Output Relation \circled{3} defines the expected follow-up value. In the example from Figure~\ref{fig:MetamorphicRuntimeLeoRover}, the Output Relation for the steering angle of the ADS consists of inverting the output from the source test case, i.e., $steering_f \simeq -steering_s$.

To obtain an uncertainty score with this Output Relation, we emit \emph{quantitative verdicts} based on the Euclidean distance, which is a simple way to measure the distance between the expected and actual values of the follow-up outputs:
\begin{equation}
    u = \sqrt{(OR(O_s) - O_f)^2}
\end{equation}
where $u$ is the uncertainty score, $OR(O_s)$ is the expected follow-up output from the model for the follow-up image, computed with the Output Relation $OR$ and the actual output from the model for the source image $O_s$, and $O_f$ is the actual output from the model for the follow-up image.

For instance, the uncertainty score for the Output Relation described above would be computed as:
\begin{equation}
    u = \sqrt{(-steering_s - steering_f)^2}
\end{equation}

Under nominal driving conditions in which the DNN is operating reliably, the MR is expected to hold within a certain error margin, and so the uncertainty scores are expected to remain under a certain boundary. However, unexpected situations that may make the ADS unreliable, such as environmental anomalies or inappropriate DNN training, could cause more severe MR violations, and thus increased uncertainty scores. This is why the verdicts emitted by \tool can be interpreted as a measure of the uncertainty for the DNN. i.e., an uncertainty score. 

Since this technique is based on the coherence between multiple inputs and output groups, there is a risk that the output for the original (source) input might be correct, and that only the follow-up output is incorrect. We argue that, as long as the follow-up input is within the validity domain for the DNN, the behavior of the DNN might still be uncertain in the case of a high deviation from the MRs, as the follow-up input is derived from (and usually similar to) the source input. For the purpose of defining MRs, the validity of an input should be determined by whether the system is expected to be able to handle such inputs in a specific way. For example, transformations that generate unrealistic images that would not be observed in the real operational environment would not be considered valid, as the DNN is not designed to handle such inputs, and therefore the MR will be prone to yielding high uncertainty scores that are not useful in practice.

Note that the described approach assumes that the driving model does not maintain some internal state that can change based on the observed inputs. In the case of Recurrent Neural Networks (RNNs) \cite{hopfield_neural_1982,rumelhart_learning_1986}, which have memory of the previous inputs, the approach would need to be adapted by saving and restoring the internal state of the models before inference. This way, two sets of internal states would be used for the driving model: (1) The state derived from the real (source) inputs, and (2) the state derived from the generated follow-up inputs.

While these quantitative verdicts can quantify the uncertainty of the DNN under test, some reference must be defined to determine whether a given value is large enough to warrant raising an alarm. This reference will differ depending on the monitoring technique and the monitored DNN, and it may even be appropriate to define different references for specific scenarios or tasks that the DNN can perform. A common strategy, which we employ in our implementation, is to define a threshold value based on the verdicts observed in nominal driving conditions, such as in (a subset of) the training data for the DNN. For instance, the uncertainty scores of some runtime monitors can be fitted into a suitable probability distribution, such that the threshold can be adjusted based on the expected false alarm rate that can be considered acceptable \cite{stocco2020misbehaviour}. However, our experiments under nominal conditions showed that, for some of the approaches, the uncertainty distributions could not be modeled with known closed-formula distributions. Therefore, in order to have a common threshold-setting method for all the approaches, \tool currently implements a simpler method which consists in using a threshold value proportional to the maximum uncertainty score observed in the training data, as described in Section \ref{sec:Evaluation:Configuration:Thresholds}.

The uncertainty score computed by an MR may be susceptible to short-term increases for a few frames, which may not be desirable to interpret as an anomalous scenario or system misbehavior. Thus, similar to SelfOracle~\cite{stocco2020misbehaviour}, we implement a time-series based analysis \circled{4} in order to smooth uncertainty score spikes and avoid false alarms due to single-frame outliers. Specifically, we implement a simple auto-regressive (AR) filter that computes the uncertainty score as a linear combination of past values. The following equation describes the computation of the uncertainty score $u_t$ at time $t$ using the last $k$ values:

\begin{equation}
    u_t = \sum^{k}_{i=1}{\frac{1}{i}u_{t-i}}
\end{equation}

\subsection{Metamorphic Relations}

We define five specific MRs for image-based ADSs, which we have implemented on top of \tool. The first four MRs define image transformations that are expected to not change the output from the DNN under nominal conditions, whereas MR5 defines a transformation which should result in a symmetric steering angle as the output from the DNN. We note that the proposed MRs are not necessarily novel. Indeed, the fourth and fifth MRs were proposed in prior studies related to metamorphic testing of image-based DNNs~\cite{spieker2020adaptive,arrieta2022multi}, whereas we did not find any study where the first three were used as MRs. In contrast, the novelty of our approach lies in the usage of MRs at runtime with the goal of measuring the uncertainty of the DNN. While our approach is generic to any kind of MRs for image-based ADSs, it is important to use MRs that are computationally cheap. We therefore used five MRs below, and avoid using others that are computationally more time consuming, such as frosting the image. The specific parameters we use for these MRs in our experiments are described in Section \ref{sec:Evaluation:Configuration:MarMot}.

\begin{itemize}
    \item \textbf{MR1: Reduce Brightness}. This transformation consists of reducing the brightness of the input image. This brightness reduction has been tuned to be within the expected tolerance of the DNN. Thus, the output is expected to be very similar to the original image.
    
    \item \textbf{MR2: Increase Contrast}. This transformation consists of increasing the contrast of the input image. This contrast increase has been tuned to be within the expected tolerance of the DNN, so the output is expected to be very similar to the original image.

    \item \textbf{MR3: Add Noise}. This transformation consists of slightly altering the values from individual pixels of the image, with a uniform distribution for the degree of change. The degree of the alterations has been tuned to be within the expected tolerance of the DNN. Thus, the output is expected to be very similar from the original image.

    \item  \textbf{MR4: Blur}. This transformation consists of slightly blurring the image. The degree of blurring has been tuned to be within the expected tolerance of the DNN. Thus, the output is expected to be very similar to the original image.

    \item \textbf{MR5: Horizontal Flip}. This transformation consists of horizontally flipping (i.e., mirroring) the image, as shown in the example from Figure~\ref{fig:MetamorphicRuntimeLeoRover}. The output from the DNN is expected to be mirrored (i.e., negated) from the one obtained with the original image. This MR assumes that the DNN uses a frontal camera mounted in the middle of the vehicle, as is the case both of our case studies.

\end{itemize}

Note that all of these MRs are designed for lane-keeping, although they may coincidentally be applicable to other tasks. For instance, MRs 1 through 4 may also be applicable to overtaking maneuvers, but MR5: Horizontal Flip would not be valid, since overtaking can usually only be done from one side.


%% file: Evaluation.tex
This section describes our empirical evaluation and presents the results obtained from our experiments. Specifically, we aim at answering the following research questions (RQs):

\begin{itemize}
    \item [\textbf{RQ1}] \textbf{Effectiveness -- }
    How effective is \tool at predicting failures of ADSs derived from internal or external anomalies? Which are the best MRs?
       
    \item [\textbf{RQ2}] \textbf{Comparison -- }
    How does \tool compare with other state-of-the-art runtime monitors?

    \item [\textbf{RQ3}]
    \textbf{Execution Cost -- }
    What is the execution cost and performance of \tool, and how does it compare to other state-of-the-art runtime monitors?
    
    \item [\textbf{RQ4}] \textbf{Reaction -- }
    How do the results for \tool and other state-of-the-art runtime monitors change as we reduce the reaction window?

    \item [\textbf{RQ5}] 
    \textbf{Frontier of Behaviors -- }
     Is \tool prone to false positives under anomalies that do not cause the system to misbehave?

\end{itemize}

\subsection{Case Study Systems}

\subsubsection{LeoRover System}

We evaluate \tool on a physical environment using the LeoRover~\cite{LeoRoverGithub} vehicle controlled by an image-based DNN provided by the vendor. Unlike previous ADS monitoring approaches~\cite{stocco2020misbehaviour,stocco2022thirdeye,hussain2022deepguard}, we employed a physical system to avoid the threats to validity produced by the reality gap between simulated and physical ADSs. While we acknowledge that our small-scale ADS is still far from a real self-driving car, it is also important to note that obtaining a dataset of anomalies and DNN faults with a real car under real conditions requires extensive resources and may potentially incur in ethical issues due to safety concerns. We therefore believe that our small scale ADS is a good step towards the assessment of monitoring techniques in real cars in more realistic conditions.

\begin{figure}[h]
    \centering
    \includegraphics[width=0.75\linewidth]{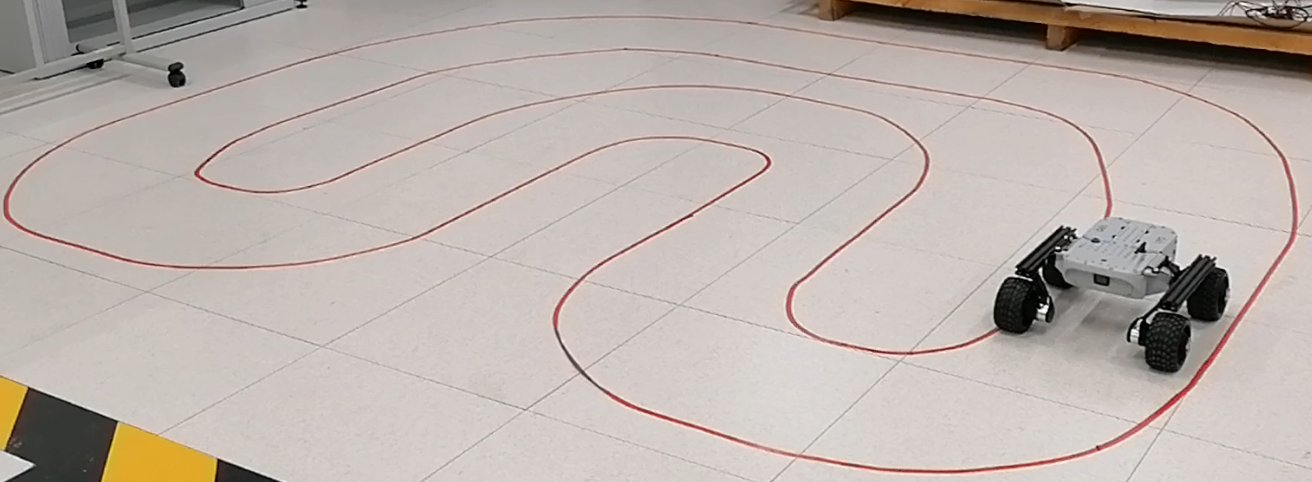}
    \vspace{-5pt}
    \caption{LeoRover in Circuit-1}
    \label{fig:Circuit-1}
\end{figure}

\begin{figure}[h]
    \centering
    \includegraphics[width=0.75\linewidth]{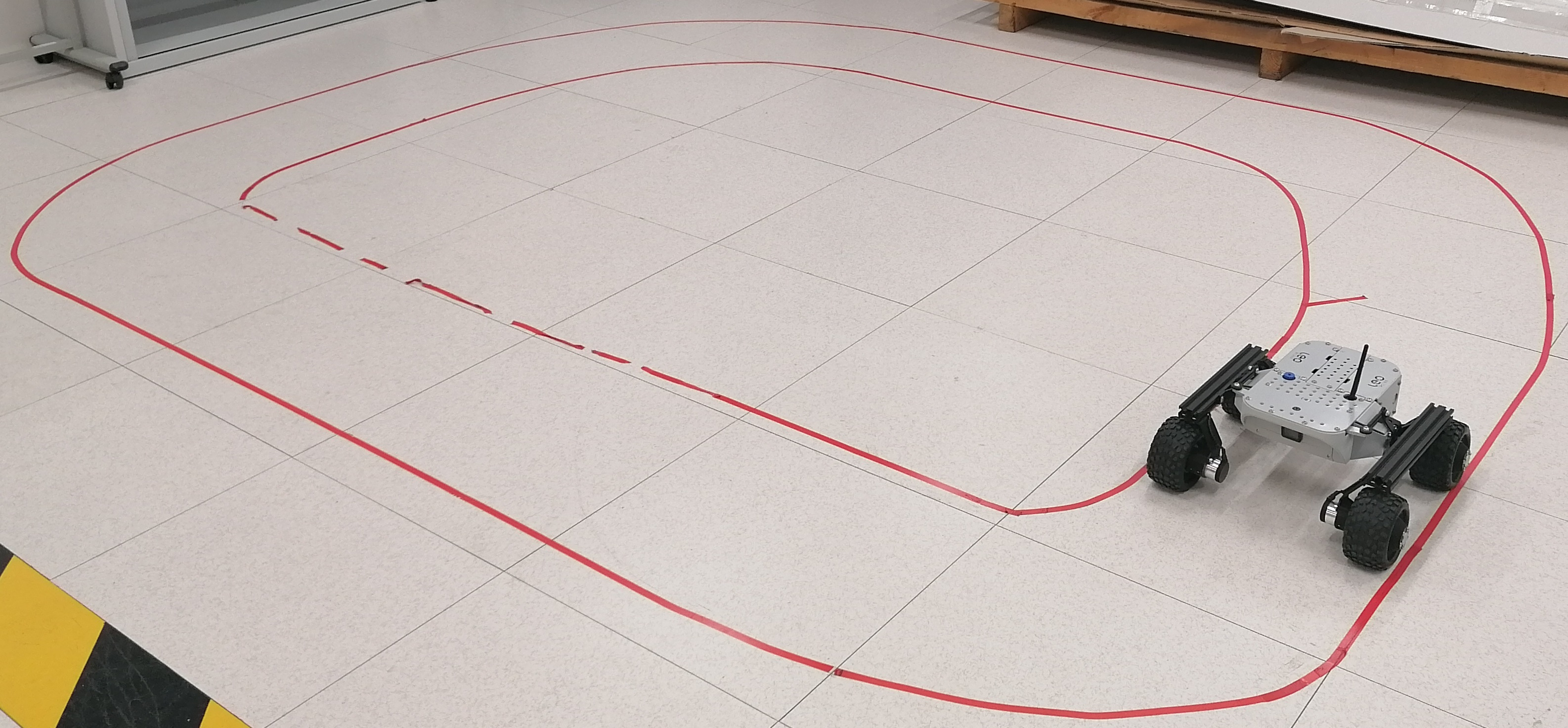}
    \vspace{-5pt}
    \caption{LeoRover in Circuit-2}
    \label{fig:Circuit-2}
\end{figure}

Our case study system's DNN is trained to follow a road delimited by colored tape. Figures \ref{fig:Circuit-1} and \ref{fig:Circuit-2} show the LeoRover vehicle driving in the two circuits we employ in our evaluation. The vehicle has a camera, which we set to record images at \textasciitilde 10 frames per second (FPS), and four static wheels with independent motors for steering and throttling, with a maximum linear speed of \textasciitilde 0.4 m/s. The DNN model consists of five convolutional layers followed by two dense layers. The DNN was trained by ourselves for 100 epochs with a combination of the dataset provided by the vendor and an additional dataset recorded by us, which we describe in Section \ref{sec:Dataset}. The software of the LeoRover employs the Robotic Operating System (ROS)~\cite{koubaa2017robot}, and the different ADS components from Figure \ref{fig:MetamorphicRuntimeLeoRover} are implemented as ROS nodes. This software architecture enables the interaction with the DNN required by \tool to be efficiently implemented via ROS messages.

\subsubsection{Nvidia’s DAVE-2 model on Udacity Simulator}

\begin{figure}[h]
    \centering
    \includegraphics[width=0.6\linewidth]{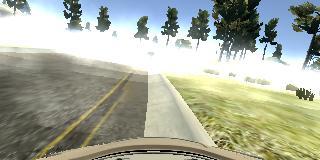}
    \vspace{-5pt}
    \caption{Dave-2 simulation out of bounds caused by fog}
    \label{fig:Dave-2-Simulation}
\end{figure}

To further validate the effectiveness and versatility of \tool, we conducted additional experiments utilizing the NVIDIA's DAVE-2 model within the Udacity Simulator environment, following the same experimental setup as Stocco et al.~\cite{stocco2022thirdeye}. The DAVE-2 model serves as the backbone of our simulated autonomous driving system (ADS). This model is renowned for its robustness and accuracy in perceiving and navigating through diverse driving scenarios. Its architecture is comprised of three convolutional layers with a 5x5 kernel and a stride of 2, alongside two additional convolutional layers with a 3x3 kernel (no stride applied). These layers are succeeded by five fully-connected layers incorporating a dropout rate of 0.05 and employing the Rectified Linear Unit (ReLU) activation function. We used the same DNN models used by Stocco et al.~\cite{stocco2022thirdeye}, which they provide in their replication package. Figure \ref{fig:Dave-2-Simulation} shows an example image recorded with the Udacity Simulator, in which the ADS is going out of bounds because of a fog anomaly.

\subsection{Datasets for LeoRover Case Study System}\label{sec:Dataset}
In this section, we describe the datasets employed to train our subject DNN and the evaluation datasets, in addition to the processes we used to collect them. To allow for a fair and reproducible empirical evaluation, we evaluated all the DNN supervisors offline. All the datasets consist of ordered lists of camera images and annotations of which frames correspond with specific events, such as out-of-bounds episodes. The images were recorded at \textasciitilde 10 FPS, i.e., 10 images roughly correspond with a second of recording. Each image is stored with a resolution of $120 \times 160$ and three color channels (RGB). Because we employed a physical ADS, significant manual effort was required for the collection of the dataset, lasting in total a 4 person-week effort. Figure \ref{fig:anomalies_tree} shows the different types of anomalies we used to generate the datasets for LeoRover. As can be seen, there are two main types of anomalies: (1) internal, which refers to anomalies in the DNN model itself, and (2) external anomalies, which refers to anomalies that occur outside the DNN model. The external anomalies are further divided into two different sub-types: (1) External Filter-based anomalies, which are generated by applying a filter to the images that the DNN receives as an input and (2) External Environmental anomalies, which are anomalies in the physical environment where the DNN is operating that could affect its driving performance.

\usetikzlibrary{trees, positioning, shapes, shadows, arrows.meta}

\tikzset{
  basic/.style  = {draw, text width=2cm, drop shadow, font=\sffamily, rectangle},
  root/.style   = {basic, rounded corners=2pt, thin, align=center, fill=white},
  level-1/.style = {basic, rounded corners=6pt, thin,align=center, fill=white, text width=3cm},
  level-2/.style = {basic, rounded corners=6pt, thin,align=center, fill=white, text width=3cm},
  level-3/.style = {basic, thin, align=center, fill=white, text width=3cm}
}

\begin{figure}[htb]
    \resizebox{.8\linewidth}{!}{
        \centering
        \begin{tikzpicture}[
            level 1/.style={sibling distance=24em, level distance=3.25em},
            level 2/.style={sibling distance=16em, level distance=3.25em},
            edge from parent/.style={->,solid,black,thick,draw}, 
            edge from parent path={(\tikzparentnode.south) -- (\tikzchildnode.north)},
            >=latex, node distance=.8cm, edge from parent fork down
        ]

        \node[root] (root) {\textbf{Anomalies}}

        child {node[level-1] (c1) {\textbf{\hyperref[subsubsec:anomalies-internal]{Internal}}}}
        child {node[level-1] (c2) {\textbf{External}}
            child {node[level-2] (c2a) {\textbf{\hyperref[subsubsec:anomalies-external-filters]{Filter-based}}}}
            child {node[level-2] (c2b) {\textbf{\hyperref[subsubsec:anomalies-external-environmental]{Environmental}}}}
        };

        \begin{scope}[every node/.style={level-3}]
        \node [below of = c1, xshift=35pt] (c1a) {Mutant HLR};
        \node [below of = c1a] (c1b) {Mutant TAN};
        \node [below of = c1b] (c1c) {Mutant TCL};

        \node [below of = c2a, xshift=35pt] (c2a1) {Gaussian noise};
        \node [below of = c2a1] (c2a2) {Shot noise};
        \node [below of = c2a2] (c2a3) {Impulse noise};
        \node [below of = c2a3] (c2a4) {Defocus blur};
        \node [below of = c2a4] (c2a5) {Fog};
        \node [below of = c2a5] (c2a6) {Brightness};
        \node [below of = c2a6] (c2a7) {Contrast};
        \node [below of = c2a7] (c2a8) {Pixelate};
        \node [below of = c2a8] (c2a9) {JPEG};
        \node [below of = c2a9] (c2a10) {Speckle noise};
        \node [below of = c2a10] (c2a11) {Gaussian blur};
        \node [below of = c2a11] (c2a12) {Spatter};
        \node [below of = c2a12] (c2a13) {Saturation};
        
        \node [below of = c2b, xshift=35pt] (c2b1) {Cover single line};
        \node [below of = c2b1] (c2b2) {Cover both lines};
        \node [below of = c2b2] (c2b3) {Flashlight};
        \node [below of = c2b3] (c2b4) {Fork};
        \node [below of = c2b4] (c2b5) {Intersection};
        \node [below of = c2b5] (c2b6) {Darkening};
        \node [below of = c2b6] (c2b7) {Obstacle};
        \end{scope}

        \foreach \value in {a, b, c}
            \draw[->] (c1.195) |- (c1\value.west);
        
        \foreach \value in {1, 2, 3, 4, 5, 6, 7, 8, 9, 10, 11, 12, 13}
            \draw[->] (c2a.195) |- (c2a\value.west);

        \foreach \value in {1, 2, 3, 4, 5, 6, 7}
        \draw[->] (c2b.195) |- (c2b\value.west);

        \end{tikzpicture}
    }
    \caption{Anomalies Tree Map}
    \label{fig:anomalies_tree}
\end{figure}
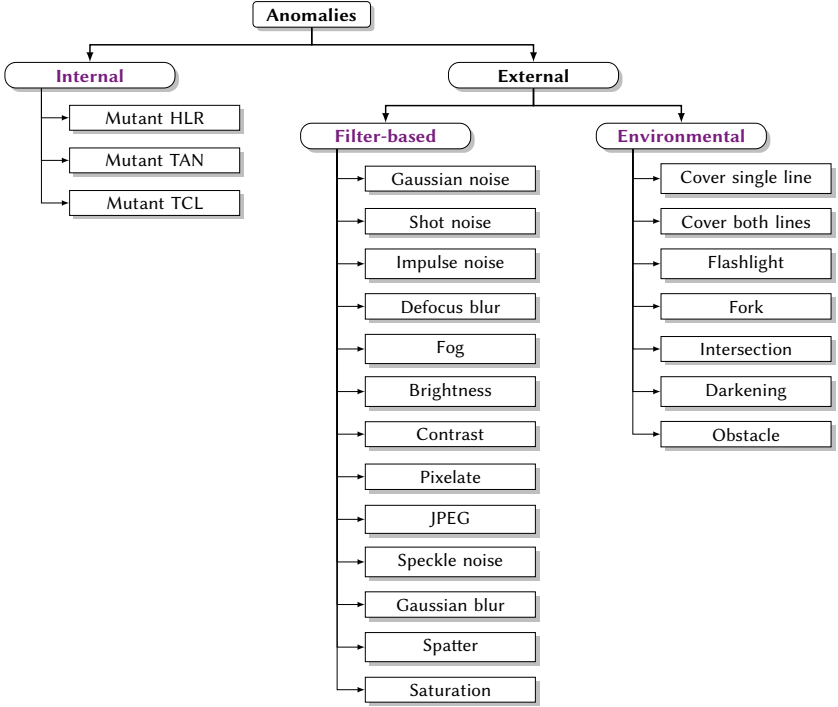

\subsubsection{Training Data} 

To train the DNN model, we combined two datasets recorded in two different circuits. The first dataset is provided by the LeoRover vendor, and consists of 10,493 labeled images. The second dataset was collected by ourselves in a test circuit, and consists of 2,832 labeled images. We extended the vendor's dataset to withstand relatively complex maneuvers (e.g., sharp curves). The circuits we used for the training were not used for evaluation purposes later. The datasets were collected by manually driving the vehicle through the circuits with a remote controller and recording the image, throttle, and steering angle for each frame. We employ different circuits for training and evaluating the model in order to ensure that the model is general enough to drive through different circuits and not overfitted to our evaluation circuits, which were both set up in the same room. Nevertheless, both datasets have a similar setting (lane-keeping, indoors lab, white floor, circuit delimited with colored tape) and were recorded with the same hardware (LeoRover).

\subsubsection{Nominal Conditions} \label{sec:Evaluation:Datasets:Nominal}

The following datasets, which we used for the evaluation of our approach and the baselines, were collected by ourselves in two circuits, different from the ones employed for training the driving DNN. These circuits, which we named Circuit-1 and Circuit-2, can be seen in Figure \ref{fig:Circuit-1} and Figure \ref{fig:Circuit-2} respectively. As it can be seen in the images, Circuit-1 contains more curves and sharper turns. In comparison, Circuit-2 only contains four curves in the same direction, but we also incorporated some additional sources of environmental uncertainty (as categorized by Zhang et al.~\cite{zhang2019uncertainty,zhang2016understanding}) in the form of discontinuous lines and an extra line inside the circuit. We verified that our DNN could successfully drive through both of these circuits under nominal conditions. Specifically, the DNN could drive 30 laps in either circuit without interruptions and without going out-of-bounds at any point.

The specific conditions we set for an out-of-bounds is both front wheels crossing the lines delimiting the circuit, such that the line can no longer be seen in the camera image. We chose this definition because it is generally an irrecoverable state, whereas a smart enough system should be able to recover if e.g. only one of the front wheels is outside, as the line is still visible.

To evaluate the rate of false alarms obtained with our approaches, we first collected recordings of the DNN driving in both circuits under nominal conditions, with no out-of-bounds events nor any other incidents. These recordings were arranged such that each of them corresponds roughly with one complete lap to the circuit.

Overall, for the nominal conditions dataset, we obtained a total of 37 recordings (13,416 images) for Circuit-1 and 35 recordings for Circuit-2 (8,320 images), each of which corresponds with one lap to the circuit.

\subsubsection{External Filter-based Anomalies Causing Misbehaviors}
\label{subsubsec:anomalies-external-filters}
To measure the ability of our DNN supervisors to identify external anomalies, we collected a dataset of out-of-bounds episodes resulting from unexpected anomalies in the images. For this purpose, we implemented the image corruptions and perturbations proposed by Hendrycks et al. \cite{hendrycks2019robustness}, which are widely used for testing image-based DNNs \cite{ferreira2021benchmarking,ruff2021unifying,guerin2022evaluation,stocco2022mind}. For our experiments, we could only use 13 out of the 19 proposed image corruptions due to performance limitations when applying the image transformations while the vehicle is running. For each of the anomalies, five different levels are employed.\footnote{The specific values for each of the anomalies can be found in \url{https://github.com/jonayerdi/marmot/blob/master/leorover_scripts/hendrycks.py}} The employed image corruptions include the following: 

\begin{itemize}
    \item Gaussian noise
    \item Shot noise
    \item Impulse noise
    \item Defocus blur
    \item Fog
    \item Brightness
    \item Contrast
    \item Pixelate
    \item JPEG compression
    \item Speckle noise
    \item Gaussian blur
    \item Spatter
    \item Saturation
\end{itemize}

To emulate the occurrence of these external anomalies, we modified the DNN script to apply the image transformation to every image to be processed by the DNN, starting after 20 seconds from the moment the first image is processed. Therefore, each recording for external anomalies consists of \textasciitilde 20 seconds of nominal driving conditions, followed by a window of anomalous external conditions which lasts until the out-of-bounds occurs. The frames in which the anomalous external conditions begin and the out-of-bounds occurs were annotated for each recording by manually inspecting the images.

Overall, for the external filter-based anomalies dataset, we obtained a total of 52 recordings (13,948 images) for Circuit-1 and other 52 recordings for Circuit-2 (12,963 images), each of which ends with an out-of-bounds episode after an image perturbation is applied. Specifically, we have 13 types of anomalies, with 2 severity parameterizations for each anomaly, and we collect 2 recordings for each anomaly type and parameterization, hence $13 \times 2 \times 2 = 52$ recordings for each circuit.

\subsubsection{Internal }Anomalies Causing Misbehaviors
\label{subsubsec:anomalies-internal}
To measure the ability of our DNN supervisors to identify internal anomalies, we collected a dataset of out-of-bounds episodes caused by faulty driving models. For this purpose, we implemented the mutation operators for DNNs proposed by Humatova et al.~\cite{humbatova2021deepcrime}, which are based on real deep learning faults. These mutation operators include inadequate training data and sub-optimal choice of the model’s architecture or of the training hyper-parameters~\cite{humbatova2021deepcrime}.

Specifically, we implemented three different mutation operators that, according to the evaluation carried out in the original paper~\cite{humbatova2021deepcrime}, were not in the group of non-killable mutants, they did not have a high triviality score and they did not produce redundant mutants. The first operator, HLR, changes the learning rate. We used two different parameterizations for this mutation, with learning rate values of  0.0001 and 0.01 (the original learning rate was 0.001). The second operator, TAN, introduces low-quality training data by adding noise to a percentage of pixels. We used two different parameterizations for this mutation, changing 15\% and 25\% of the pixels from the images. Lastly, TCL, changes labels of training data, mimicking wrong labeling situations. Again for this case, two different parameterizations were used, assigning a random value to the steering angle of the vehicle to 15\% and 20\% of the images. For each mutant and parameterization, we train 10 different models, since the mutants are stochastic~\cite{humbatova2021deepcrime}.

For this dataset, we collected recordings of the mutant DNNs driving through the circuits until an out-of-bounds episode occurred. The frames in which the out-of-bounds occur were annotated for each recording by manually inspecting the images.

Overall, for the internal anomalies dataset, we obtained a total of 30 recordings (13,339 images) for Circuit-1 and other 30 recordings for Circuit-2 (6,689 images), each of which ends with an out-of-bounds episode. Specifically, we used three types of mutations, with two parameterizations for each mutation, and we collected five recordings for each mutant type and parameterization, hence $3 \times 2 \times 5 = 30$ recordings for each circuit. For each recording, we randomly selected one of the 10 models we generated for the respective mutation operator and parameterization, meaning that not all the mutant models were used for collecting this dataset.

It is interesting to note that, for this dataset, there is a large discrepancy between both circuits in the number of images recorded. As we  will later mention in Section \ref{sec:Evaluation:Reaction:LeoRover}, in Circuit-1 the mutants misbehave (go out-of-bounds) within 40 seconds from the beginning of the recording on average, whereas the average time until a misbehavior occurs in Circuit-2 is 20 seconds. Since we stop recording images shortly after a misbehavior occurs, the average length of the recordings for Circuit-2 is correspondingly around half of Circuit-1. We hypothesize that the extra environmental anomalies we introduced in Circuit-2, namely, the discontinuous lines and the extra line inside the circuit (see Figure \ref{fig:Circuit-2}), might cause the mutated DNNs to go out-of-bounds faster. The more complex layout of Circuit-1 may also cause the LeoRover to turn sharply more often, slowing its movement and thus increasing the time until the out-of-bounds occurs.

\subsubsection{External Environmental }Anomalies Causing Misbehaviors
\label{subsubsec:anomalies-external-environmental}
We evaluated the capability of \tool and chosen baseline methods to detect external environmental anomalies. To achieve this, we compiled a dataset comprising out-of-bounds episodes caused by unforeseen anomalies in the rover's environment. We introduced a total of seven real environmental anomalies that could be encountered by any driver on the road to facilitate this evaluation.

\begin{itemize}
    \item Cover single line
    \item Cover both lines
    \item Flashlight
    \item Fork
    \item Intersection
    \item Darkening
    \item Obstacle
\end{itemize}

To simulate the occurrence of external environmental anomalies, we manually introduced these anomalies (e.g., an obstacle placed on one side of the road as the rover passed). Consequently, all recordings comprised several seconds of nominal driving conditions, followed by a period of anomalous external environmental circumstances, culminating in an out-of-bounds event. Similar to prior recordings, we manually annotated the frames marking the onset of the anomalous external environmental circumstances and the moment the out-of-bounds event occurred by reviewing the images.

For the external environmental anomalies, we completed a total of 9 recordings, encompassing 1,245 images for Circuit-1 and an additional 9 recordings with 1,283 images for Circuit-2. Each recording concludes with an out-of-bounds episode after introducing an an external environmental anomaly in the rover's path. Among the seven types of external environmental anomalies we used, we deemed it particularly insightful to document the flashlight and obstacle anomalies across both straight paths and curves. These anomalies exhibited the most distinct behaviors between curves and straights. Conversely, the remaining external environmental anomalies showed similar behaviors in both scenarios, negating the need for distinct recordings for curves and straights.

\subsubsection{Non-misbehaving Anomalies} \label{sec:Dataset:Passing}

To evaluate the effectiveness of \tool under the presence of anomalies that do not cause the system to misbehave, we collected datasets with internal, external filter-based, and external environmental anomalies with no out-of-bounds episodes. Each recording of this dataset consists of roughly a full lap to the circuit. The inputs from these recordings should be valid, but close to the frontier of behaviors for the DNN, i.e., closer to provoking misbehaviors than nominal conditions.

Overall, for the external filter-based anomalies dataset, we obtained a total of 26 recordings (5,176 images) for Circuit-1 and other 26 recordings for Circuit-2 (3,842 images). For this dataset, we employed a less severe parameterization for the anomaly filters, such that the DNN would still be able to successfully complete laps in both circuits. Specifically, we have 13 types of anomalies, with a single severity parameterization for each anomaly, and we collect 2 recordings for each anomaly type and parameterization, hence $13 \times 2 = 26$ recordings for each circuit.

For the external environmental anomalies dataset, we obtained a total of 9 recordings (1,393 images) for Circuit-1 and other 9 recordings for Circuit-2 (1,610 images). We employed the same types of anomalies as for the equivalent dataset with out-of-bounds episodes, but with less intense disturbances.

For the internal anomalies dataset, we obtained a total of 30 recordings (27,447 images) for Circuit-1 and other 30 recordings for Circuit-2 (20,149 images). We employ the same three mutation operators as for the dataset with misbehaviors, but with less severe parameterizations, such that the DNN would still be able to successfully complete laps in both circuits. Specifically, we used three types of mutations, with two parameterizations for each mutation, and we collected five recordings for each mutant type and parameterization, hence $3 \times 2 \times 5 = 30$ recordings for each circuit.

\subsection{Datasets for NVIDIA's Dave-2 Model}

We used the same dataset as Stocco et al., which is provided with their replication package~\cite{stocco2022thirdeye}. 
Similar to the LeoRover case study system, to allow for a fair comparison, we evaluated all the DNN supervisors offline. All the datasets consist of ordered lists of camera images and annotations of which frames correspond with out-of-bounds episodes. The images were recorded at \textasciitilde 10 FPS, i.e., 10 images roughly correspond with a second of recording. Each image is stored with a resolution of $320 \times 160$ and three color channels (RGB).

\subsubsection{Training Data}
The replication package provided by Stocco et al. \cite{stocco2022thirdeye} contains a recording with 16,542 images of driving through the training track under nominal conditions, and another recording with 15,696 images driving the same track in the reverse direction under nominal conditions. This is the dataset from Track 1 (Lake) recorded by Stocco et al. refer to their evaluation runs as  \cite{stocco2020misbehaviour}. We used the scripts from their replication package in order to train all the Dave-2 models with this dataset for our experiments.

\subsubsection{Nominal Conditions}
To compute false positives and true negatives in the evaluation of our approach and the baselines, we employed the ``normal'' recording of Track 1 (Lake) from the training dataset, which consists of 16,542 images of driving through the training track under nominal conditions.

\subsubsection{External Anomalies Causing Misbehaviors}
The dataset of anomalies causing misbehaviors (i.e., out-of-bounds episodes) consists of 31 recordings in which adverse weather effects are simulated. Specifically, 30 of the recordings simulate rain, fog or snow with 10 different varying degrees of intensity in the range $[10\%, 20\%, ..., 100\%]$, thus, 10 recordings for each weather condition: (10 $\times$ rain, 10 $\times$ fog, 10 $\times$ snow) \cite{stocco2022thirdeye}. The final recording simulates day/night cycles, which is a non-parameterizable effect \cite{stocco2022thirdeye}.

Each recording contains \emph{multiple} annotated out-of-bounds episodes. Our evaluation considers whether each of these out-of-bounds episodes is detected by the oracles separately for true positives and false negatives (each out-of-bounds should be detected during the in-bounds period right before). To avoid detection periods that are too short, we require that the period during which the vehicle is in-bounds before going out-of-bounds must be at least 100 frames long (roughly 3 or 4 seconds), otherwise the out-of-bounds episode is ignored in our evaluation. With this criteria, there are a total of 65 individual out-of-bounds episodes that can be detected by DNN runtime monitors in this dataset.

\subsubsection{Internal Anomalies Causing Misbehaviors}
The dataset of internal anomalies causing misbehaviors consists of 18 recordings in which faulty DNNs were used to drive the vehicle \cite{stocco2022thirdeye}. More specifically, each of these recordings consists of one lap of Track 1 driven by a mutant of the Dave-2 model generated with the DeepCrime mutation testing tool~\cite{humbatova2021deepcrime}. The 18 different recordings use different mutants, with unique mutation operators and/or parameterizations.

Each recording contains \emph{multiple} annotated out-of-bounds episodes. Same as with the anomalies dataset, we considered each of these out-of-bounds episodes individually, as long as the in-bounds period before is at least 100 frames long. In total, there are 180 individual out-of-bounds episodes that can be detected by the runtime monitors in this dataset.

\subsection{Baselines}

\subsubsection{SelfOracle}

To evaluate the effectiveness of \tool, we compared the obtained results with SelfOracle~\cite{stocco2020misbehaviour}, a state-of-the-art misbehaviour prediction technique based on input image reconstruction models. SelfOracle~\cite{stocco2020misbehaviour} estimates the uncertainty for the DNN based on the reconstruction error of the driving model's input images. During training, an autoencoder model is trained with the same images used to train the subject DNN. At runtime, the input images are reconstructed by the autoencoder model, and the uncertainty score is measured as the reconstruction error.

\begin{equation}
    u = \frac{1}{WHC}\sum^{W,H,C}_{i=1,j=1,c=1}{(x[c][i,j] - x'[c][i,j])^2}
\end{equation}

\noindent where $u$ is the uncertainty score, $x_t$ is the input image, $x'$ is the image reconstructed by the autoencoder from $x$, and $W,H,C$ are the pixel-wise width, height, and channels of the input images. The rationale behind this is that images with a high reconstruction error may be out of the distribution from the training data, and therefore, the DNNs behavior may be unreliable for those inputs.

We specifically employed 4 different autoencoder models implemented in their open-source release\footnote{https://github.com/testingautomated-usi/selforacle} for our experiments, namely: (1) SAE (simple autoencoder with a single hidden layer), (2) DAE (deep autoencoder with five fully-connected layers), (3) CAE (convolutional autoencoder with alternating convolutional and max-pooling layers), and (4) VAE (variational autoencoder). We could not train the LSTM variant due to the limited availability of computational resources. Note, however, that the VAE was the best configuration of SelfOracle~\cite{stocco2020misbehaviour}, and was also used in other similar evaluations~\cite{stocco2022thirdeye,hussain2022deepguard}.

These autoencoder models have been trained with the images from the datasets used to train the driving model, since this is the approach that would most likely be used if SelfOracle was to be adopted in practice.

\subsubsection{Ensemble} \label{sec:Evaluation:Baselines:Ensemble}

Furthermore, we also employed a simple monitor based on Deep Ensemble Neural Networks \cite{lakshminarayanan_simple_2017} as an additional baseline. Our Ensemble monitor employs 10 driving DNNs trained with different weight initializations and training/validation splits of the dataset, but the same architecture and hyperparameters. The ensemble prediction is calculated as the average of the outputs of the individual driving models for each input image. Similarly, the uncertainty score for that monitor is computed as the standard deviation of the outputs. Thus, we can obtain the uncertainty score using this technique as follows:

\begin{equation}\label{eq:std}
   u = \sqrt{\frac{1}{M}\sum^{M}_{i=1}{\left(O_i - \left(\frac{1}{M}\sum^{M}_{j=1}{O_j}\right)\right)^2}}
\end{equation}

\noindent where $u$ is the uncertainty score, $M$ is the number of models used in the ensemble, and $O_i$ is the output from model $i$.

The computation of the standard deviation elucidates the variance in model predictions, acting as an indicator of uncertainty. Higher standard deviation values indicate greater disagreement among the models, potentially highlighting inputs that are more difficult or ambiguous for the DNNs to interpret. In practice, this method does not require major architectural changes to the network and has proven to be a good predictor of uncertainty~\cite{weiss2023uncertainty}. However, having multiple models is computationally expensive at both training and inference time.

\subsubsection{MC Dropout} \label{sec:Evaluation:Baselines:MC_Dropout}
\begin{figure}[h]
    \centering
    \includegraphics[width=0.75\linewidth]{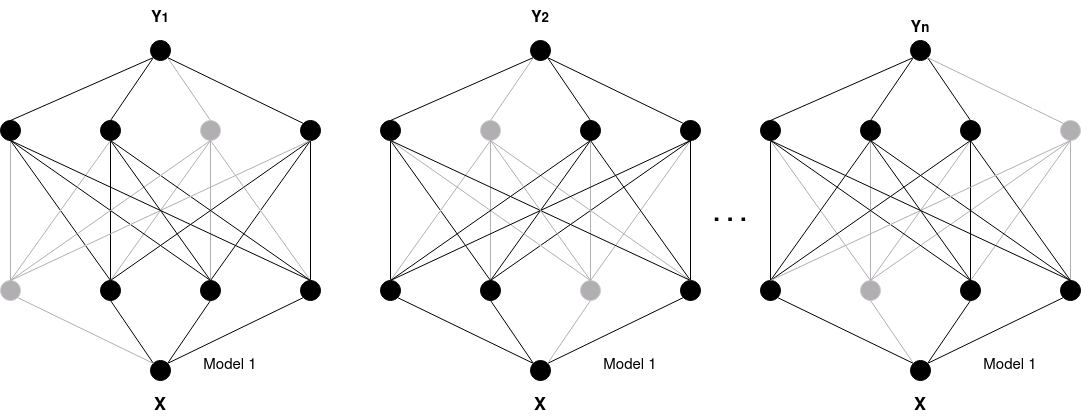}
    \vspace{-5pt}
    \caption{MC Dropout representation}
    \label{fig:MCDropout}
\end{figure}

MC Dropout is a technique that randomly deactivates a certain percentage of neurons in the DNN (not only at training time but also during inference) to create an ensemble effect with a single model \cite{gal2016MCDropout}. This approach allows the uncertainty of the model to be estimated by performing multiple forward passes (see Figure \ref{fig:MCDropout}), resulting in multiple stochastic predictions, and obtaining their standard deviation as shown in Equation~\ref{eq:std}.

A higher standard deviation indicates greater uncertainty in the model's predictions, signaling areas where the model may lack information or where the data may be inherently ambiguous. The key advantage of MC Dropout is its simplicity in approximating Bayesian inference in neural networks, eliminating the need for intricate training procedures. On the other hand, it implies having at least one dropout layer in the network architecture and requires a similar computational effort as Deep Ensembles at inference time.


\subsection{Configuration}

\subsubsection{Thresholds} \label{sec:Evaluation:Configuration:Thresholds}
As mentioned at the end of Section \ref{sec:Approach:Architecture}, the uncertainty scores produced by \tool require a threshold value to determine whether the uncertainty score indicates normal conditions or is too high and an alarm needs to be triggered. For this purpose, we isolated 7 out of the 37 recordings of nominal driving conditions from Circuit-1, which correspond with 7 laps to the circuit (The uncertainty distributions observed in this experiment can be found in Appendix \ref{apx:uncertaintydist}). These 7 recordings were used as a reference to determine the threshold value for each of our oracles, using the following formula:

\begin{equation}
    t_o = max\left(max\left(u_o(f) \; \forall \; f \in R\right) \; \forall \; R \in D\right) \cdot 1.1
\end{equation}

\noindent where $t_o$ is the threshold for the oracle $o$, $max\left(u_o(f) \; \forall \; f \in R\right))$ is the maximum uncertainty score obtained with the oracle $o$ from all the frames $f$ in recording $R$, and $D$ is the set of all recordings used to compute the thresholds. We chose to add a margin of 10\% for computing the threshold in order to avoid false alarms, hence the $1.1$ multiplier. We consider that the oracle $o$ raises an alarm at any point when its uncertainty score is greater than $t_o$.

We considered individual MRs from \tool and individual autoencoder models from SelfOracle to be distinct oracles, so we computed thresholds for each of them separately.

We discarded the recordings used for computing these thresholds from our evaluation dataset. Thus, our evaluation dataset for nominal conditions consists of 30 recordings from Circuit-1 and all the 35 recordings from Circuit-2.

\subsubsection{Time-Series Analysis}

For all runtime monitors, we applied the simple auto-regressive filter described in Section \ref{sec:approach} using the last 10 images ($k = 10$). Note that the implementation of this filter is equivalent to the one proposed in SelfOracle~\cite{stocco2020misbehaviour}, so we do not deviate from their proposed approach.

\subsubsection{\tool} \label{sec:Evaluation:Configuration:MarMot}
We used the following parameters for the implementation of our MRs. We ensured that the follow-up inputs generated with the selected MR parameters followed the input validity of the domain as defined by Riccio and Tonella~\cite{riccio2023and}. Specifically, we performed a manual inspection of the transformed images to ensure that the road visibility was still good, and we also checked that the DNNs could drive correctly with the transformed inputs under nominal conditions.

\begin{itemize}
    \item \emph{MR1: Reduce Brightness.} We substract 77 from the values of each pixel (saturating substraction, so values remain between 0 and 255).
    \item \emph{MR2: Increase Contrast.} We convert the image to HSV format and set the saturation value of every pixel to 50 (possible values range from 0 to 255).

    \item \emph{MR3: Add Noise.} We alter the values of each pixel by a rate from 0\% to 20\% (uniformly sampled random value per pixel).

    \item \emph{MR4: Blur.} We apply a normalized box filter, as implemented by OpenCV-Python's \texttt{cv2.blur} function, using a kernel size of $1 \times 5$.

\end{itemize}

\subsubsection{SelfOracle}
For training the SelfOracle autoencoder models, image augmentation was applied using flipping, translation, shadowing, and brightness changes. 60\% of the data was augmented through these transformations, matching the configuration used by Stocco et al.~\cite{stocco2020misbehaviour}. It is noteworthy that this algorithm is also employed in DeepGuard~\cite{hussain2022deepguard}.

Due to the stochastic nature of training these autoencoder models, we trained each model 10 times with different random seeds, which affects the initialization and the image augmentations applied. All the results for SelfOracle reported in Section~\ref{sec:Evaluation:Results} are average values for the 10 models of each autoencoder type.

\subsubsection{Ensemble}
As mentioned in Section \ref{sec:Evaluation:Baselines:Ensemble}, the Ensemble oracle employs 10 different driving DNNs trained with different initializations. For each of the mutations of the driving DNN employed for our internal anomalies dataset, we also trained 10 models with different initializations, and the Ensemble oracle employs the corresponding mutants of the driving models when evaluating the internal anomalies dataset.

\subsubsection{MC Dropout}
For each driving DNN model, including mutants mimicking internal anomalies, we trained a corresponding model with modified dropout layers that have an effect at inference time. These stochastic models are used by the MC Dropout oracle when evaluating driving DNNs. For our LeoRover driving model, we employ a dropout rate of 0.1 in the dropout layers, which are already interleaved between all layers in the original model. For the Dave-2 driving model, we used the training scripts provided with the replication packages from Thirdeye~\cite{stocco2022thirdeye} and Deepcrime~\cite{humbatova2021deepcrime}, which interleave dropout layers with dropout rates of 0.05 between all layers for MC Dropout. In all cases, we perform 16 inferences in order to compute uncertainty scores with this approach.

\subsection{Evaluation Metrics}

For each recording described in Section \ref{sec:Dataset}, a test oracle may or may not raise an alarm if its uncertainty score is greater than its threshold (Section \ref{sec:Evaluation:Configuration:Thresholds}). For each recording from the nominal conditions dataset, we define an instance where an alarm is raised as a false positive ($FP$), and an instance where it is not as a true negative ($TN$). Conversely, for each recording from the internal or external anomalies datasets, we define an instance where an alarm is raised as a true positive ($TP$), and an instance where it is not as a false negative ($FN$).

In our evaluation, we only considered whether an oracle fired an alarm or not for each recording, since we assumed that the alarms might cause a response and consecutive alarms will have no effect. Therefore, every individual recording described in Section \ref{sec:Dataset} will count as exactly one FP or TN (nominal conditions dataset), or one TP or FN (anomalies or mutants datasets).

Similar to most evaluations using classifiers, we also report the false positive rate ($FPR = \frac{FP}{FP+TN}$) for the nominal conditions dataset, as well as the precision ($Prec. = \frac{TP}{TP+FP}$), the recall or true positive rate ($TPR = \frac{TP}{TP+FN}$), and the F1-score ($F1 = 2 \cdot \frac{Prec. \cdot TPR}{Prec. + TPR}$) for the anomalies datasets.

Similar to Stocco et al.~\cite{stocco2020misbehaviour}, we also report two threshold-independent metrics for evaluating classifiers: AUC-ROC (area under the curve of the Receiver Operating Characteristics) and AUC-PRC (area under the Precision-Recall Curve).

In addition to the accuracy metrics, evaluating the performance efficiency of different proposed methods is crucial, especially for their application in resource-constrained environments such as the LeoRover. For RQ3, we specifically assess the execution time per image, system CPU usage, and system RAM consumption of each runtime monitor, as these factors directly impact the feasibility and reliability of deploying these techniques in real-world scenarios.

\section{Analysis of the Results and Discussion} \label{sec:Evaluation:Results}

\subsection{RQ1 -- Effectiveness}

\subsubsection{LeoRover Dataset} \phantom \\

\label{sec:Evaluation:Reaction:LeoRover}

Column ``Nominal'' from Table  \ref{table:ResultsLeoRoverThresholds} shows the evaluation results for the nominal datasets (i.e., there is no misbehavior) from Circuit-1 and Circuit-2. With our chosen threshold, for \tool, we observe three FPs for MR5 (Horizontal Flip) in Circuit-2, which corresponds with an FPR (to be minimized) of 10\%. For all other cases, we observe 0 FPs for all of \tool's MRs.

Column ``External Filter Anomaly'' from Table \ref{table:ResultsLeoRoverThresholds} shows the evaluation results for the external filter-based anomaly datasets with misbehaviors (i.e., out-of-bounds episodes). In terms of TPR (the higher, the better), we observe that MR5 achieves the best results in both circuits, 35\% for Circuit-1 and 65\% for Circuit-2. In comparison, the rest of the MRs achieve significantly lower TPR and F1 measures, indicating a much lower ability to identify external filter-based anomalies. 

On the other hand, column ``External Environ. Anomaly'' from Table \ref{table:ResultsLeoRoverThresholds} shows the results for the external environmental anomalies, which were implemented by physically manipulating the environment in which the vehicle operates. In this case, we also observe that MR5 achieves the best results in both circuits in terms of both TPR (44\% and 56\% for circuits 1 and 2 respectively) and F1 scores (62\% and 59\% for circuits 1 and 2 respectively).

Finally, column ``Internal Anomaly'' from Table \ref{table:ResultsLeoRoverThresholds} shows the evaluation results for the internal anomaly datasets (i.e., faults of the DNN based on the three chosen mutation operators from Deepcrime~\cite{humbatova2021deepcrime}). In this case, we observe that all of \tool's MRs achieve very high TPR and F1 scores, with TPRs higher than 77\% in all cases. Furthermore, MR5 achieves 100\% TPR in both circuits.

\input{tables_results}
Table \ref{table:ResultsLeoRoverAUC} shows the threshold-independent measures of AUC-PRC and AUC-ROC for all the anomaly datasets with misbehaviors. For both external filter-based and internal anomalies, we observe that MR5 (Horizontal Flip) achieves the best results in all cases, with MR1 (Reduce Brightness) and MR3 (Add Noise) being the next best MRs. Surprisingly, although MR5 is still highly effective in the external environmental anomaly datasets, MR1 achieves the best results in Circuit-1, whereas MR2 (Increase Contrast) achieves the best results in Circuit-2.

The reason why the effectiveness of \tool is much lower for identifying external anomalies compared with internal anomalies might be related with the length of the reaction period, i.e., the period between the beginning of the anomalous scenario until the moment where the vehicle goes out-of-bounds. We observed that most of the image filters we implemented for the anomalies dataset cause the vehicle to drive off the road fairly quickly, leaving a monitor with a very short period to react. Conversely, many of the mutant DNNs are capable of driving within bounds for longer periods of time, which gives the monitors a much longer time to react to the DNNs uncertain behavior. The estimated average reaction period for the anomalies datasets recordings is around 7 seconds for Circuit-1 and 5 seconds for Circuit-2, whereas the reaction periods for the mutants datasets are around 40 seconds for Circuit-1 and 20 seconds for Circuit-2.

\subsubsection{Nvidia's Dave2 Dataset} \phantom \\

\input{tables_results_dave2}

Column ``Nominal'' from Table  \ref{table:ResultsDave2Thresholds} shows the evaluation results for the nominal datasets (i.e., there is no misbehavior) from the NVIDIA's Dave-2 case study system with the Udacity simulator. With our chosen threshold for \tool, we observe a single FP (FPR of 10\%) for MR1 and MR4, and two FPs (FPR of 20\%) for MR2 and MR5, whereas MR4 did not trigger FPs. On the other hand, Column ``External Anomaly'' from Table  \ref{table:ResultsDave2Thresholds} shows that MR5 (Horizontal Flip) achieves the highest TPR (54\%) and F1 score (69\%) from all the \tool MRs. Furthermore, MR1 (Reduce brightness) and MR3 (Add Noise) also appear to obtain significant TPR (34\%) and F1 scores (50\% and 51\%).

As for Column ``Internal Anomaly'' from Table \ref{table:ResultsDave2Thresholds}, MR2 (Increase Contrast) achieved the highest TPR of 88\% and F1 score of 93\%. Furthermore, MR1 and MR3 also appear to obtain significant TPR (64\%) and F1 scores (78\%). Table~\ref{table:ResultsDave2AUC} shows the threshold-independent measures of AUC-PRC and AUC-ROC for all the datasets with internal and external anomalies for the NVIDIA's Dave-2 case study system. These results roughly match the ones obtained with thresholds in Table \ref{table:ResultsDave2Thresholds}, with MR5 obtaining the highest AUC-ROC (75.3\%) for external anomalies and MR2 achieving the highest AUC-ROC (93.8\%) for internal anomalies. It is noteworthy that the AUC-ROC of MR2 for internal anomalies is much higher than the other MRs, which do not reach 60\% in any case. We also observe that, for the external anomalies, MR1 and MR3 achieve an AUC-PRC score of 91.5\%, which is slightly higher than the one obtained by MR5 (90.9\%).

\subsubsection*{Concluding Remark for RQ1} \phantom \\

Based on the evaluation of our approach on two different case study systems and datasets, we can answer the first RQ as follows:

\begin{custombox}{RQ1}
In summary, we can conclude that \tool is effective at identifying internal anomalies, with TPRs of 100\% in the LeoRover case study and 88\% in the Dave-2 case study, while also being fairly effective at identifying external anomalies, with the best TPRs ranging from 35\% to 65\% in the LeoRover case study and 54\% in the Dave-2 case study.
\end{custombox}

\subsection{RQ2 -- Comparison}
\subsubsection{LeoRover Dataset} \phantom \\

Column ``Nominal'' from Table \ref{table:ResultsLeoRoverThresholds} shows that all oracles but Ensemble achieve 0 FPs for Circuit-1, whereas for Circuit-2, Ensemble, SelfOracle CAE (single FP in only one of the 10 executions) and MR5 (Horizontal Flip) are the only oracles with FPs. Overall, the FPR (to be minimized) is not higher than 10\% for any of the oracles, but the highest FPR is obtained by \tool's MR5 in Circuit-2. Column ``External Filter Anomaly'' from Table \ref{table:ResultsLeoRoverThresholds} shows that, in terms of TPR and F1 (to be maximized), MR5 achieves the best results in both circuits, followed by Ensemble, and then either MR1 (Reduce Brightness) or SelfOracle with the CAE model. Furthermore, we also observe that MR5 achieves the best TPR and F1 in both circuits in Column ``External Environ. Anomaly''. Lastly, we observe from column ``Internal Anomaly'' from Table \ref{table:ResultsLeoRoverThresholds} that all of \tool's MRs achieve a significantly higher TPR and F1 score than the other approaches, followed by Ensemble. Here, SelfOracle achieves TPRs of 12\% or lower with CAE in both circuits, and 0\% scores with the other autoencoder models. It is worth noting that the MC Dropout baseline obtains scores of 0 across the board for both external filter-based and external environmental anomalies, and only appears to be effective at identifying misbehaviors from internal anomalies.

Based on the threshold-independent measures from Table \ref{table:ResultsLeoRoverAUC}, we observe that all of \tool's MRs outperform the SelfOracle, Ensemble and MC Dropout baselines by a significant margin for the internal anomaly datasets. Regarding the external environmental anomalies, \tool also outperforms all baselines with any of the MRs in almost all cases. As for the external filter-based anomalies, MR5 achieves the highest scores in Circuit-1, closely followed by SelfOracle's CAE model, and then followed by Ensemble. Surprisingly, Ensemble achieves the highest scores for the Circuit-2 anomalies, followed by \tool's MRs, and then followed by SelfOracle's CAE.

Comparing the results from Circuit-1 and Circuit-2 for the anomalies datasets, we observe that all monitors obtained noticeably better results in Circuit-2, despite the fact that the reaction period for this circuit is slightly shorter on average (\textasciitilde7 seconds for Circuit-1 and \textasciitilde5 seconds for Circuit-2). We speculate that the extra environmental uncertainties introduced in Circuit-2 (i.e., the discontinuous road boundaries and the extra line in the middle of the road) causes a slightly more erratic driving, which might make the uncertain behavior from anomalies to become more noticeable for the monitors. The results from Table \ref{table:ResultsLeoRoverAUC} seem to indicate that Ensemble and MC Dropout are the most favored by this factor. This may be due to the fact that both of them are white-box approaches, and thus potentially more sensitive to the DNNs internal erratic behavior than \tool and SelfOracle, which are black-box. 

\subsubsection{Nvidia's Dave2 Dataset} \phantom \\

Column ``Nominal'' from Table \ref{table:ResultsDave2Thresholds} shows that \tool's MR1 (Reduce Brightness) and MR4 (Blur) obtain a single FP (FPR of 10\%), while SelfOracle's VAE obtains 1.7 FPs on average (0-4 FPs depending on the run, average FPR of 17\%). Lastly, \tool's MR2 (Increase Contrast) and MR5 (Horizontal Flip), as well as MC Dropout, obtain two FPs each (FPR of 20\%).

Regarding Column ``External Anomaly'' from Table \ref{table:ResultsDave2Thresholds}, we can see that SelfOracle's VAE achieves the best results by a significant margin, with a TPR of 100\% and an F1 score of 99\%. \tool's MR5 obtains the second best results, with a TPR of 54\% and an F1 score of 69\%. On the other hand, Column ``Internal Anomaly'' shows that \tool's MR2 achieves the best results, with a TPR of 88\% and a F1 score of 93\%. SelfOracle's VAE obtains the second best results, with a TPR of 79\% and a F1 score of 88\%.

A similar conclusion can be drawn from Table \ref{table:ResultsDave2AUC}, with SelfOracle's VAE achieving the highest AUC-PRC (98.4\%) and AUC-ROC (90\%) in Column ``External Anomaly'', and \tool's MR2 achieving the highest AUC-PRC (98.5\%) and AUC-ROC (93.8\%) in Column ``Internal Anomaly''.

\subsubsection*{Concluding Remark for RQ2} \phantom \\

Based on the evaluation of our approach on two different case study systems and datasets, we can answer the second RQ as follows:

\begin{custombox}{RQ2}
In summary, we can conclude that \tool outperforms all baselines in terms of identifying internal anomalies in both of our case study systems, while also achieving a comparable or higher effectiveness at identifying external anomalies with the best MRs in most cases.
\end{custombox}

\subsection{RQ3 -- Execution Cost}

\begin{figure*}[ht]
\centering
\subfloat[System CPU usage]{\vspace{-6pt}\label{fig:ExecCPU}\includegraphics[width=0.7\textwidth]{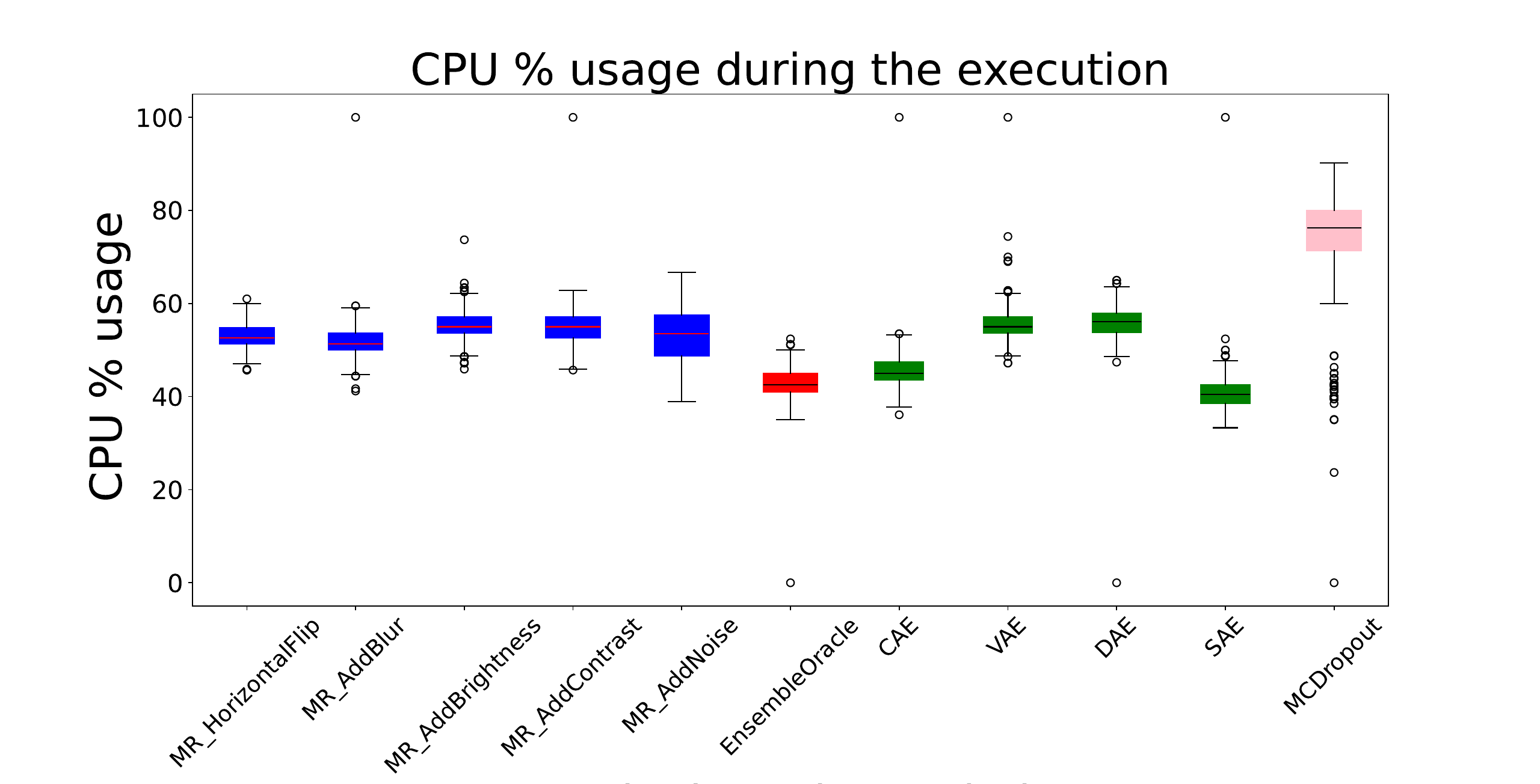}}
\,
\subfloat[System RAM usage]{\vspace{-6pt}\label{fig:ExecRAM}\includegraphics[width=0.7\textwidth]{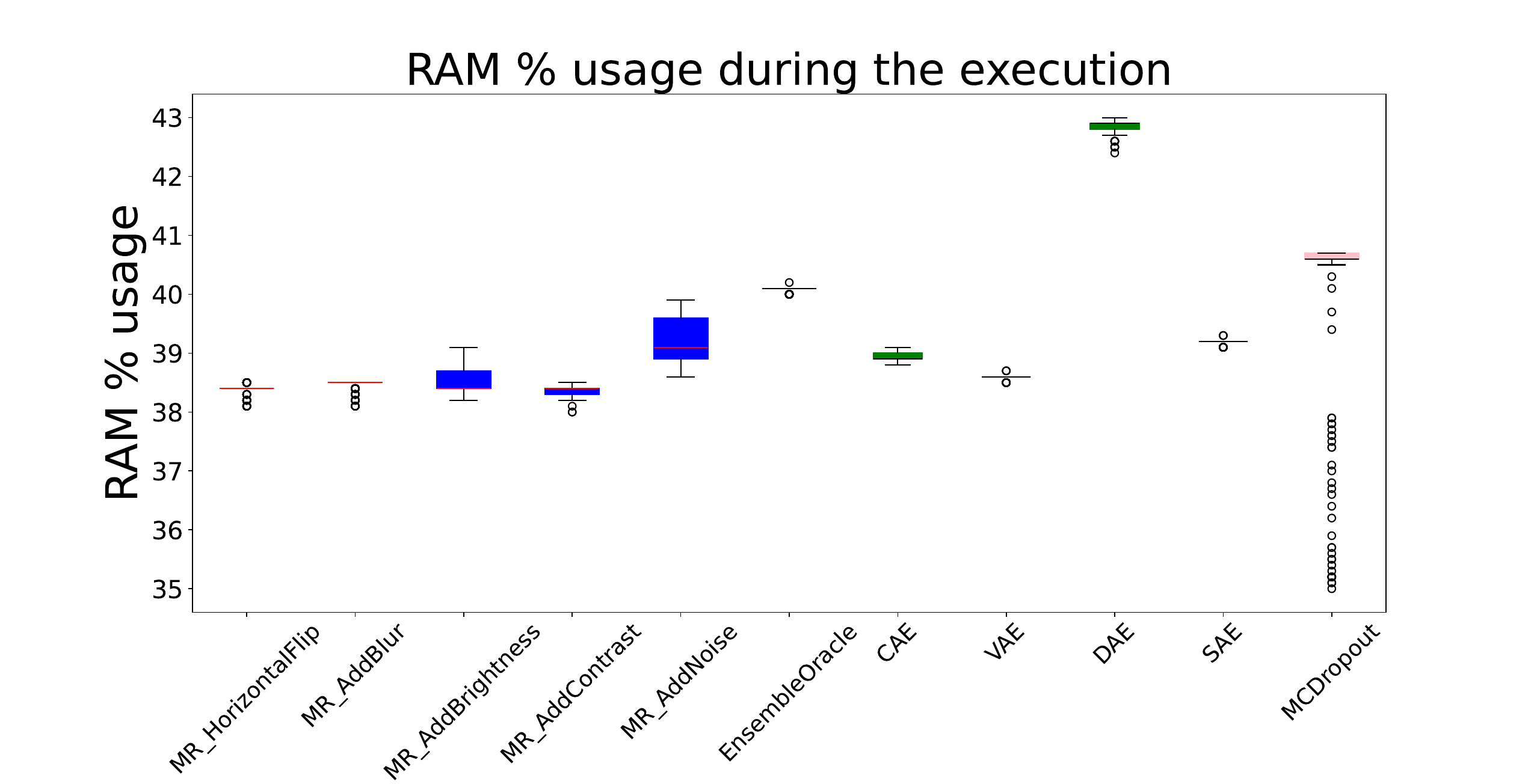}}
\,
\subfloat[Execution Time for processing a single image]{\vspace{-6pt}\label{fig:ExecTIME}\includegraphics[width=0.7\textwidth]{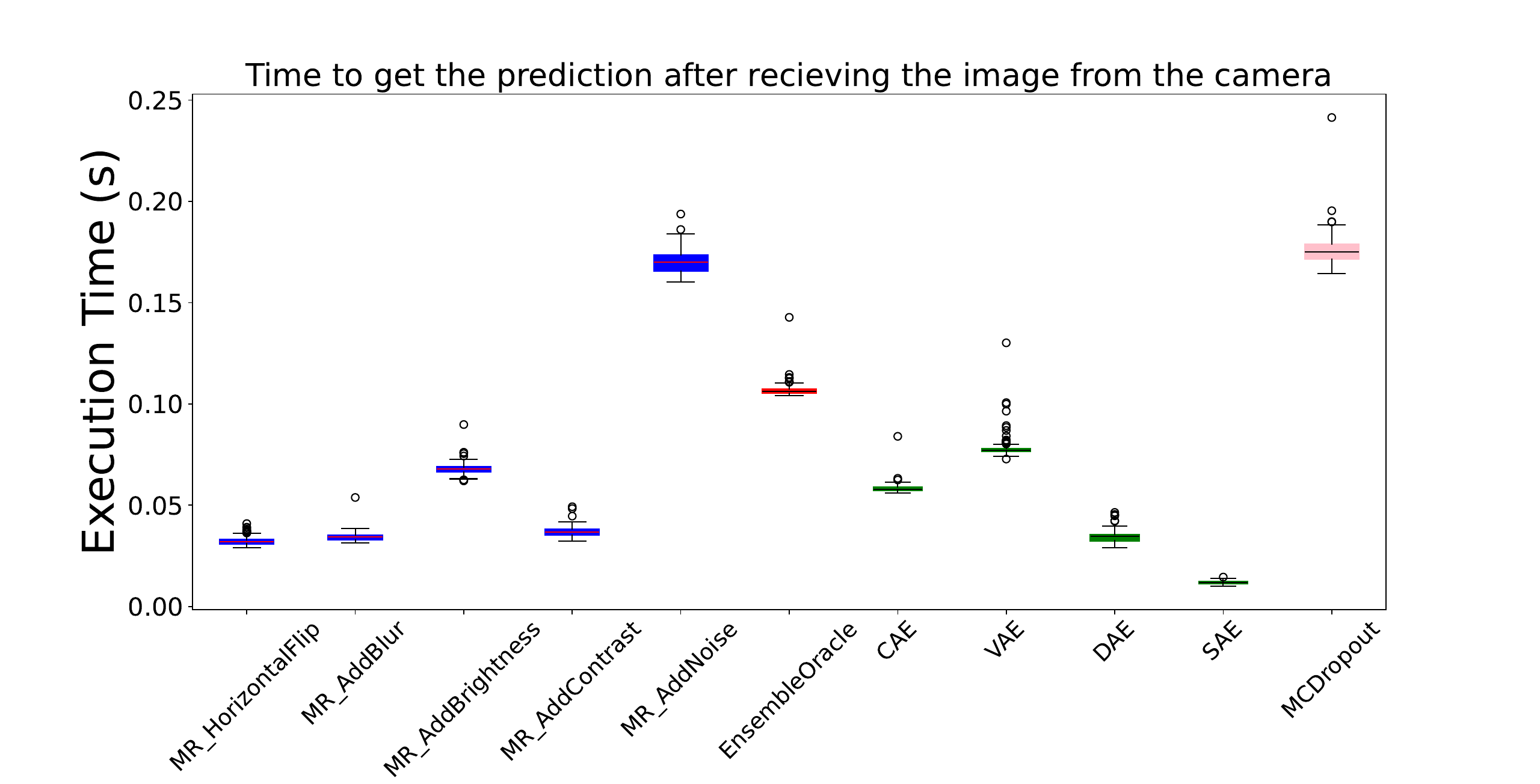}}
\vspace{-5pt}
\caption{System resource consumption when running on the rover. In blue the results for Metamorphic Relations, in red the results for  Ensemble oracle, in green the results for SelfOracles and in pink the results for MC Dropout }
\label{fig:ExecResults}
\end{figure*}

\begin{table}[h!]
\setlength{\tabcolsep}{2pt}
\renewcommand{\arraystretch}{0.7}
\caption{Comparison of monitors (Effect size for CPU usage)}
\vspace{-6pt}
\label{tab:Stats_cpu}
\resizebox{.7\textwidth}{!}{
\begin{tabular}{lccccccccccc}
\toprule
\textbf{} & \textbf{MR1} & \textbf{MR2} & \textbf{MR3} & \textbf{MR4} & \textbf{MR5} & \textbf{CAE} & \textbf{DAE} & \textbf{SAE} & \textbf{VAE} & \textbf{Ens} & \textbf{MCD} \\ \hline
\textbf{MR1} & \multicolumn{1}{c}{\cellcolor[HTML]{EEEEEE}} & - & S\textsubscript{2} & L\textsubscript{2} & M\textsubscript{2} & L\textsubscript{2} & S\textsubscript{1} & L\textsubscript{2} & - & L\textsubscript{2} & L\textsubscript{1} \\
\textbf{MR2} & - & \multicolumn{1}{c}{\cellcolor[HTML]{EEEEEE}} & S\textsubscript{2} & L\textsubscript{2} & M\textsubscript{2} & L\textsubscript{2} & S\textsubscript{1} & L\textsubscript{2} & - & L\textsubscript{2} & L\textsubscript{1} \\
\textbf{MR3} & S\textsubscript{1} & S\textsubscript{1} & \multicolumn{1}{c}{\cellcolor[HTML]{EEEEEE}} & N & - & L\textsubscript{2} & S\textsubscript{1} & L\textsubscript{2} & S\textsubscript{1} & L\textsubscript{2} & L\textsubscript{1} \\
\textbf{MR4} & L\textsubscript{1} & L\textsubscript{1} & N & \multicolumn{1}{c}{\cellcolor[HTML]{EEEEEE}} & S\textsubscript{1} & L\textsubscript{2} & L\textsubscript{1} & L\textsubscript{2} & L\textsubscript{1} & L\textsubscript{2} & L\textsubscript{1} \\
\textbf{MR5} & M\textsubscript{1} & M\textsubscript{1} & - & S\textsubscript{2} & \multicolumn{1}{c}{\cellcolor[HTML]{EEEEEE}} & L\textsubscript{2} & L\textsubscript{1} & L\textsubscript{2} & M\textsubscript{1} & L\textsubscript{2} & L\textsubscript{1} \\
\textbf{CAE} & L\textsubscript{1} & L\textsubscript{1} & L\textsubscript{1} & L\textsubscript{1} & L\textsubscript{1} & \multicolumn{1}{c}{\cellcolor[HTML]{EEEEEE}} & L\textsubscript{1} & L\textsubscript{2} & L\textsubscript{1} & M\textsubscript{2} & L\textsubscript{1} \\
\textbf{DAE} & S\textsubscript{2} & S\textsubscript{2} & S\textsubscript{2} & L\textsubscript{2} & L\textsubscript{2} & L\textsubscript{2} & \multicolumn{1}{c}{\cellcolor[HTML]{EEEEEE}} & L\textsubscript{2} & S\textsubscript{2} & L\textsubscript{2} & L\textsubscript{1} \\
\textbf{SAE} & L\textsubscript{1} & L\textsubscript{1} & L\textsubscript{1} & L\textsubscript{1} & L\textsubscript{1} & L\textsubscript{1} & L\textsubscript{1} & \multicolumn{1}{c}{\cellcolor[HTML]{EEEEEE}} & L\textsubscript{1} & M\textsubscript{1} & L\textsubscript{1} \\
\textbf{VAE} & - & - & S\textsubscript{2} & L\textsubscript{2} & M\textsubscript{2} & L\textsubscript{2} & S\textsubscript{1} & L\textsubscript{2} & \multicolumn{1}{c}{\cellcolor[HTML]{EEEEEE}} & L\textsubscript{2} & L\textsubscript{1} \\
\textbf{Ens} & L\textsubscript{1} & L\textsubscript{1} & L\textsubscript{1} & L\textsubscript{1} & L\textsubscript{1} & M\textsubscript{1} & L\textsubscript{1} & M\textsubscript{2} & L\textsubscript{1} & \multicolumn{1}{c}{\cellcolor[HTML]{EEEEEE}} & L\textsubscript{1} \\
\textbf{MCD} & L\textsubscript{2} & L\textsubscript{2} & L\textsubscript{2} & L\textsubscript{2} & L\textsubscript{2} & L\textsubscript{2} & L\textsubscript{2} & L\textsubscript{2} & L\textsubscript{2} & L\textsubscript{2} & \multicolumn{1}{c}{\cellcolor[HTML]{EEEEEE}} \\ \bottomrule
\end{tabular}}
\end{table}

\begin{table}[h!]
\setlength{\tabcolsep}{2pt}
\renewcommand{\arraystretch}{0.7}
\caption{Comparison of monitors (Effect size for RAM usage)}
\vspace{-6pt}
\label{tab:Stats_ram}
\resizebox{.7\textwidth}{!}{
\begin{tabular}{lccccccccccc}
\toprule
\textbf{} & \multicolumn{1}{l}{\textbf{MR1}} & \multicolumn{1}{l}{\textbf{MR2}} & \multicolumn{1}{l}{\textbf{MR3}} & \multicolumn{1}{l}{\textbf{MR4}} & \multicolumn{1}{l}{\textbf{MR5}} & \multicolumn{1}{l}{\textbf{CAE}} & \multicolumn{1}{l}{\textbf{DAE}} & \multicolumn{1}{l}{\textbf{SAE}} & \multicolumn{1}{l}{\textbf{VAE}} & \multicolumn{1}{l}{\textbf{Ens}} & \multicolumn{1}{l}{\textbf{MCD}} \\ \hline
\textbf{MR1} & \cellcolor[HTML]{EEEEEE} & M\textsubscript{2} & L\textsubscript{1} & N & S\textsubscript{2} & L\textsubscript{1} & L\textsubscript{1} & L\textsubscript{1} & M\textsubscript{1} & L\textsubscript{1} & L\textsubscript{1} \\
\textbf{MR2} & M\textsubscript{1} & \cellcolor[HTML]{EEEEEE} & L\textsubscript{1} & L\textsubscript{1} & S\textsubscript{1} & L\textsubscript{1} & L\textsubscript{1} & L\textsubscript{1} & L\textsubscript{1} & L\textsubscript{1} & L\textsubscript{1} \\
\textbf{MR3} & L\textsubscript{2} & L\textsubscript{2} & \cellcolor[HTML]{EEEEEE} & L\textsubscript{2} & L\textsubscript{2} & S\textsubscript{2} & L\textsubscript{1} & - & L\textsubscript{2} & L\textsubscript{1} & L\textsubscript{1} \\
\textbf{MR4} & N & L\textsubscript{2} & L\textsubscript{1} & \cellcolor[HTML]{EEEEEE} & L\textsubscript{2} & L\textsubscript{1} & L\textsubscript{1} & L\textsubscript{1} & L\textsubscript{1} & L\textsubscript{1} & L\textsubscript{1} \\
\textbf{MR5} & S\textsubscript{1} & S\textsubscript{2} & L\textsubscript{1} & L\textsubscript{1} & \cellcolor[HTML]{EEEEEE} & L\textsubscript{1} & L\textsubscript{1} & L\textsubscript{1} & L\textsubscript{1} & L\textsubscript{1} & L\textsubscript{1} \\
\textbf{CAE} & L\textsubscript{2} & L\textsubscript{2} & S\textsubscript{1} & L\textsubscript{2} & L\textsubscript{2} & \cellcolor[HTML]{EEEEEE} & L\textsubscript{1} & L\textsubscript{1} & L\textsubscript{2} & L\textsubscript{1} & L\textsubscript{1} \\
\textbf{DAE} & L\textsubscript{2} & L\textsubscript{2} & L\textsubscript{2} & L\textsubscript{2} & L\textsubscript{2} & L\textsubscript{2} & \cellcolor[HTML]{EEEEEE} & L\textsubscript{2} & L\textsubscript{2} & L\textsubscript{2} & L\textsubscript{2} \\
\textbf{SAE} & L\textsubscript{2} & L\textsubscript{2} & - & L\textsubscript{2} & L\textsubscript{2} & L\textsubscript{2} & L\textsubscript{1} & \cellcolor[HTML]{EEEEEE} & L\textsubscript{2} & L\textsubscript{1} & L\textsubscript{1} \\
\textbf{VAE} & M\textsubscript{2} & L\textsubscript{2} & L\textsubscript{1} & L\textsubscript{2} & L\textsubscript{2} & L\textsubscript{1} & L\textsubscript{1} & L\textsubscript{1} & \cellcolor[HTML]{EEEEEE} & L\textsubscript{1} & L\textsubscript{1} \\
\textbf{Ens} & L\textsubscript{2} & L\textsubscript{2} & L\textsubscript{2} & L\textsubscript{2} & L\textsubscript{2} & L\textsubscript{2} & L\textsubscript{1} & L\textsubscript{2} & L\textsubscript{2} & \cellcolor[HTML]{EEEEEE} & L\textsubscript{1} \\
\textbf{MCD} & L\textsubscript{2} & L\textsubscript{2} & L\textsubscript{2} & L\textsubscript{2} & L\textsubscript{2} & L\textsubscript{2} & L\textsubscript{1} & L\textsubscript{2} & L\textsubscript{2} & L\textsubscript{2} & \cellcolor[HTML]{EEEEEE} \\ \bottomrule
\end{tabular}}
\end{table}

\begin{table}[h!]
\setlength{\tabcolsep}{2pt}
\renewcommand{\arraystretch}{0.7}
\caption{Comparison of monitors (Effect size for execution time per image)}
\vspace{-6pt}
\label{tab:Stats_time}
\resizebox{.7\textwidth}{!}{
\begin{tabular}{lccccccccccc}
\toprule
\textbf{} & \multicolumn{1}{l}{\textbf{MR1}} & \multicolumn{1}{l}{\textbf{MR2}} & \multicolumn{1}{l}{\textbf{MR3}} & \multicolumn{1}{l}{\textbf{MR4}} & \multicolumn{1}{l}{\textbf{MR5}} & \multicolumn{1}{l}{\textbf{CAE}} & \multicolumn{1}{l}{\textbf{DAE}} & \multicolumn{1}{l}{\textbf{SAE}} & \multicolumn{1}{l}{\textbf{VAE}} & \multicolumn{1}{l}{\textbf{Ens}} & \multicolumn{1}{l}{\textbf{MCD}} \\ \hline
\textbf{MR1} & \cellcolor[HTML]{EEEEEE} & L\textsubscript{2} & L\textsubscript{1} & L\textsubscript{2} & L\textsubscript{2} & L\textsubscript{2} & L\textsubscript{2} & L\textsubscript{2} & L\textsubscript{1} & L\textsubscript{1} & L\textsubscript{1} \\
\textbf{MR2} & L\textsubscript{1} & \cellcolor[HTML]{EEEEEE} & L\textsubscript{1} & L\textsubscript{2} & L\textsubscript{2} & L\textsubscript{1} & L\textsubscript{2} & L\textsubscript{2} & L\textsubscript{1} & L\textsubscript{1} & L\textsubscript{1} \\
\textbf{MR3} & L\textsubscript{2} & L\textsubscript{2} & \cellcolor[HTML]{EEEEEE} & L\textsubscript{2} & L\textsubscript{2} & L\textsubscript{2} & L\textsubscript{2} & L\textsubscript{2} & L\textsubscript{2} & L\textsubscript{2} & L\textsubscript{1} \\
\textbf{MR4} & L\textsubscript{1} & L\textsubscript{1} & L\textsubscript{1} & \cellcolor[HTML]{EEEEEE} & L\textsubscript{2} & L\textsubscript{1} & - & L\textsubscript{2} & L\textsubscript{1} & L\textsubscript{1} & L\textsubscript{1} \\
\textbf{MR5} & L\textsubscript{1} & L\textsubscript{1} & L\textsubscript{1} & L\textsubscript{1} & \cellcolor[HTML]{EEEEEE} & L\textsubscript{1} & L\textsubscript{1} & L\textsubscript{2} & L\textsubscript{1} & L\textsubscript{1} & L\textsubscript{1} \\
\textbf{CAE} & L\textsubscript{1} & L\textsubscript{2} & L\textsubscript{1} & L\textsubscript{2} & L\textsubscript{2} & \cellcolor[HTML]{EEEEEE} & L\textsubscript{2} & L\textsubscript{2} & L\textsubscript{1} & L\textsubscript{1} & L\textsubscript{1} \\
\textbf{DAE} & L\textsubscript{1} & L\textsubscript{1} & L\textsubscript{1} & - & L\textsubscript{2} & L\textsubscript{1} & \cellcolor[HTML]{EEEEEE} & L\textsubscript{2} & L\textsubscript{1} & L\textsubscript{1} & L\textsubscript{1} \\
\textbf{SAE} & L\textsubscript{1} & L\textsubscript{1} & L\textsubscript{1} & L\textsubscript{1} & L\textsubscript{1} & L\textsubscript{1} & L\textsubscript{1} & \cellcolor[HTML]{EEEEEE} & L\textsubscript{1} & L\textsubscript{1} & L\textsubscript{1} \\
\textbf{VAE} & L\textsubscript{2} & L\textsubscript{2} & L\textsubscript{1} & L\textsubscript{2} & L\textsubscript{2} & L\textsubscript{2} & L\textsubscript{2} & L\textsubscript{2} & \cellcolor[HTML]{EEEEEE} & L\textsubscript{1} & L\textsubscript{1} \\
\textbf{Ens} & L\textsubscript{2} & L\textsubscript{2} & L\textsubscript{1} & L\textsubscript{2} & L\textsubscript{2} & L\textsubscript{2} & L\textsubscript{2} & L\textsubscript{2} & L\textsubscript{2} & \cellcolor[HTML]{EEEEEE} & L\textsubscript{1} \\
\textbf{MCD} & L\textsubscript{2} & L\textsubscript{2} & L\textsubscript{2} & L\textsubscript{2} & L\textsubscript{2} & L\textsubscript{2} & L\textsubscript{2} & L\textsubscript{2} & L\textsubscript{2} & L\textsubscript{2} & \cellcolor[HTML]{EEEEEE} \\ \bottomrule
\end{tabular}}
\end{table}

Figure \ref{fig:ExecResults} summarizes the results from the resources consumption perspective of applying \tool and all the other runtime monitoring techniques directly in the LeoRover to complete an entire lap. Specifically, we measured the total CPU usage (Figure \ref{fig:ExecCPU}), RAM usage (Figure \ref{fig:ExecRAM}) and execution time (Figure \ref{fig:ExecTIME}) for processing one image, and we sampled these values for every camera image processed. Note that the system uses the TensorFlow Lite \footnote{\url{https://www.tensorflow.org/lite}} runtime to run the models, so there is no GPU involved.

On the other hand, Tables \ref{tab:Stats_cpu}, \ref{tab:Stats_ram} and \ref{tab:Stats_time} show the results from the statistical comparisons between all monitors.
We first analyzed the data distribution using the Shapiro-Wilk test. Since the data was not normally distributed, we employed the Wilcoxon Rank-Sum test. A p-value below 0.05 was considered as significant distinction between the compared monitors, and we evaluated the effect sizes through the Vargha and Delaney \^{A}$_{12}$ value. As suggested by Romano et al.~\cite{Romano2006}, we categorized the effect size as \textbf{N} (\textit{negligible}) if $d< 0.147$, \textbf{S} (\textit{small}) if  $d<0.33$, \textbf{M} (\textit{medium}) if  $d < 0.474$ and \textbf{L} (\textit{large}) if $d >=0.474$, where $d = 2|$\^{A}$_{12}$ $-0.5|$. A subscript of \textbf{1} indicates that the metric values are smaller for the monitors in the first column, whereas a subscript of \textbf{2} indicates the opposite. Comparisons where there was no statistical significance (p-value $\geq 0.05$) are omitted.

Regarding the CPU usage, as can be seen in Figure \ref{fig:ExecCPU} and Table \ref{tab:Stats_cpu}, MC Dropout was noticeably more costly than the other approaches, with an average consumption of 74\%. Conversely, the Ensemble method and Selforacle's SAE and CAE were the ones with less CPU consumption, with average values of 40-45\%. The MRs proposed in \tool obtained similar CPU usage values as SelfOracle's VAE and DAE autoencoders, averaging CPU usages of 52-55\%.

In terms of RAM consumption, as can be seen in Figure \ref{fig:ExecRAM} and Table \ref{tab:Stats_ram}, SelfOracle's DAE method stands out for requiring the most RAM with a mean value of 43\%. Regarding the remaining methods, the Ensemble and MC dropout show a higher RAM consumption compared with the other methods: 40\% for the Ensemble method and 41\% for the MC dropout method. Nevertheless, this difference in the RAM consumption is minimal.

Finally, as shown in Figure \ref{fig:ExecTIME} and Table \ref{tab:Stats_time}, the MC Dropout and the MR3: Add Noise are the slowest methods to complete their processes, with a mean execution time of 0.175 seconds for MC Dropout and 0.169 seconds for MR3 (Add Noise). Among the other techniques, the SefOracle's SAE approach stands out for its quick execution time, with a mean value of 0.011 seconds. The rest of the approaches scored a similar execution time. The slower speed of MR3 could be attributed to our implementation, which first creates a noise array with the same size as the image and then combines it, instead of modifying the image directly. This would also explain the slightly higher RAM consumption of this MR compared with the rest.

Considering that we process the camera images at 10 FPS in our evaluation, every approach with an execution time of under 0.1 seconds should be possible to integrate without disrupting the functionality of the ADS. Only \tool's MR3, Ensemble and MC Dropout had a longer execution time, with Ensemble having a slightly higher execution time and MR3 and MC Dropout having almost twice as much.

\subsubsection*{Concluding Remark for RQ3} \phantom \\

Based on the evaluation of the different approaches onboard the LeoRover, we can answer the third RQ as follows:

\begin{custombox}{RQ3}
In summary, we can conclude that the cost of executing \tool is acceptable, and that it would be feasible to run our LeoRover experiments online, excluding a single MR which seemingly had an implementation issue. SelfOracle was found to have a similar performance, whereas Ensemble and MC Dropout were found to be slower.
\end{custombox}

\subsection{RQ4 -- Reaction}

In this RQ, we aim to account for the latency of the anomaly detection and reaction process that would exist in practice (communications latency, processing time, etc.), which might prevent the triggered healing process (e.g., stopping the vehicle before out-of-bounds) from taking effect in time. For this RQ, we consider all the datasets with misbehaviors from both case study systems.

\subsubsection{LeoRover Dataset} \phantom \\

\begin{figure*}[ht]
\centering
\subfloat[External Filter  Anomalies Circuit-1]{\label{fig:ReactionAUCPRC:AnomalyCircuit-1}\includegraphics[width=0.48\textwidth]{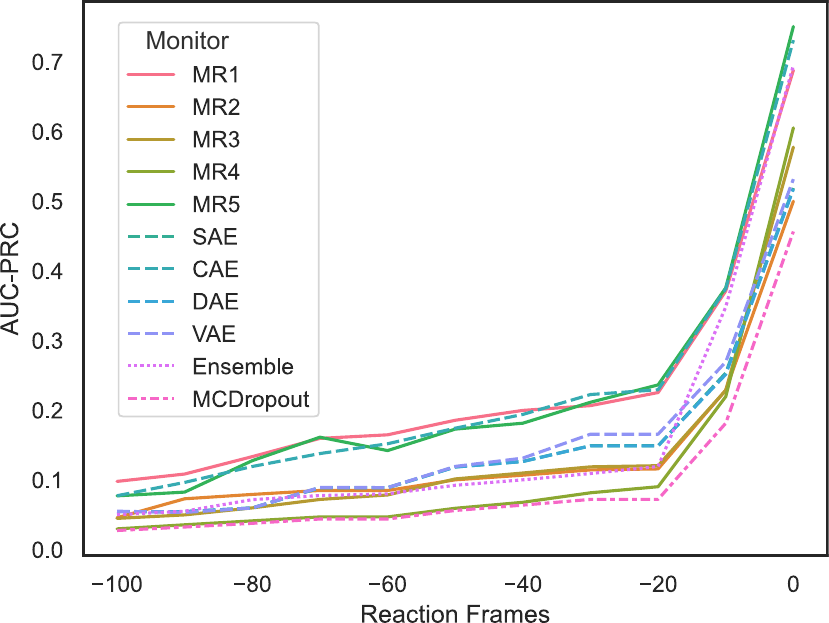}}
\,
\subfloat[External Filter Anomalies Circuit-2]{\label{fig:ReactionAUCPRC:AnomalyCircuit-2}\includegraphics[width=0.48\textwidth]{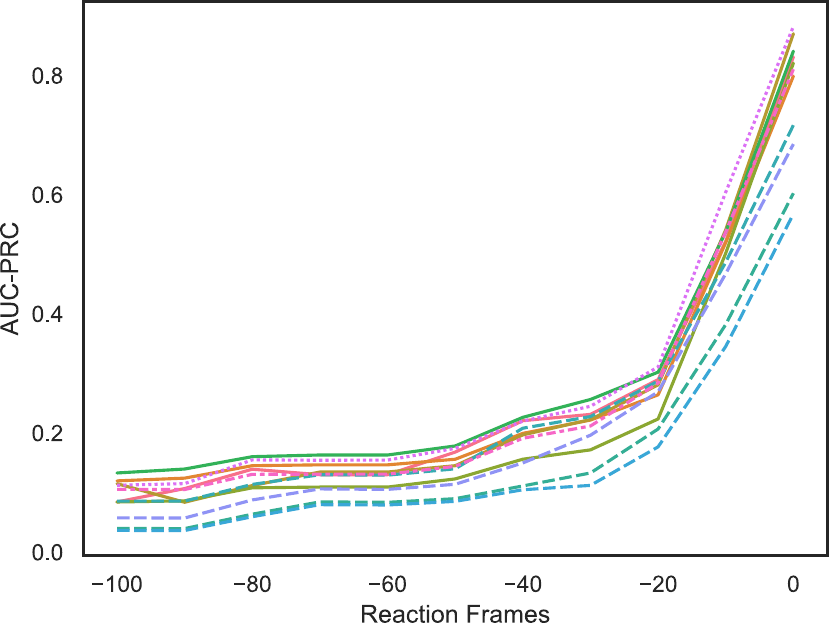}}
\,
\subfloat[External Environ. Anomalies Circuit-1]{\label{fig:ReactionAUCPRC:EnvironmentalAnomalyCircuit-1}\includegraphics[width=0.48\textwidth]{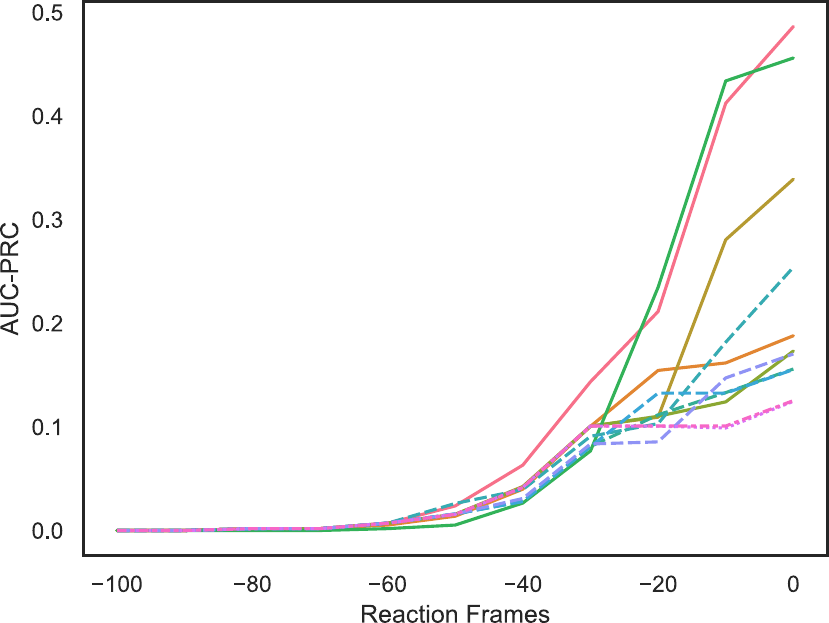}}
\,
\subfloat[External Environ. Anomalies Circuit-2]{\label{fig:ReactionAUCPRC:EnvironmentalAnomalyCircuit-2}\includegraphics[width=0.48\textwidth]{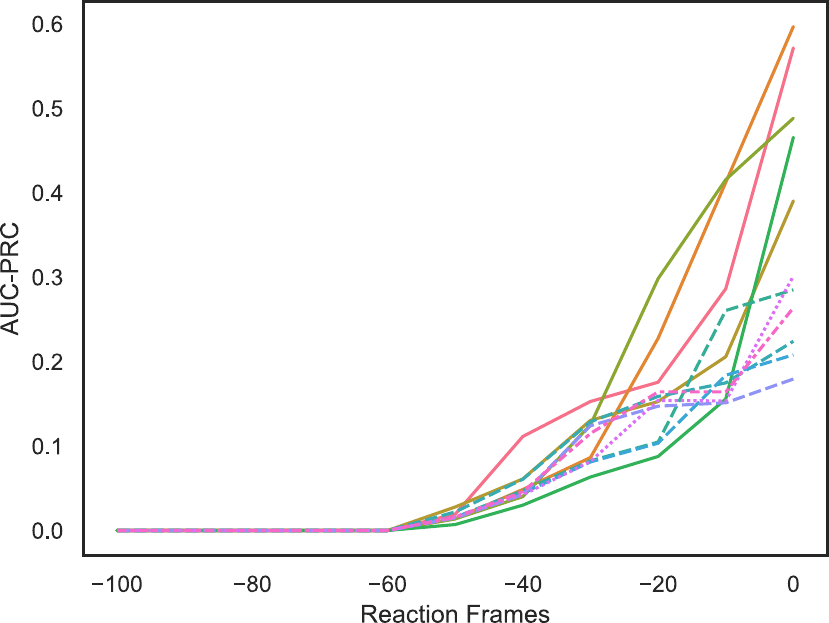}}
\,
\subfloat[Internal Anomalies Circuit-1]{\label{fig:ReactionAUCPRC:MutantCircuit-1}\includegraphics[width=0.48\textwidth]{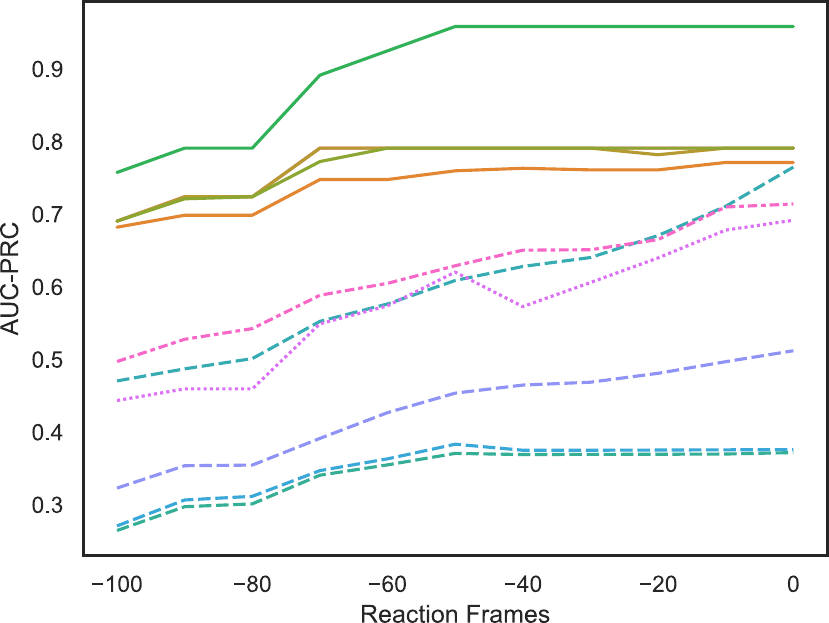}}
\,
\subfloat[Internal Anomalies Circuit-2]{\label{fig:ReactionAUCPRC:MutantCircuit-2}\includegraphics[width=0.48\textwidth]{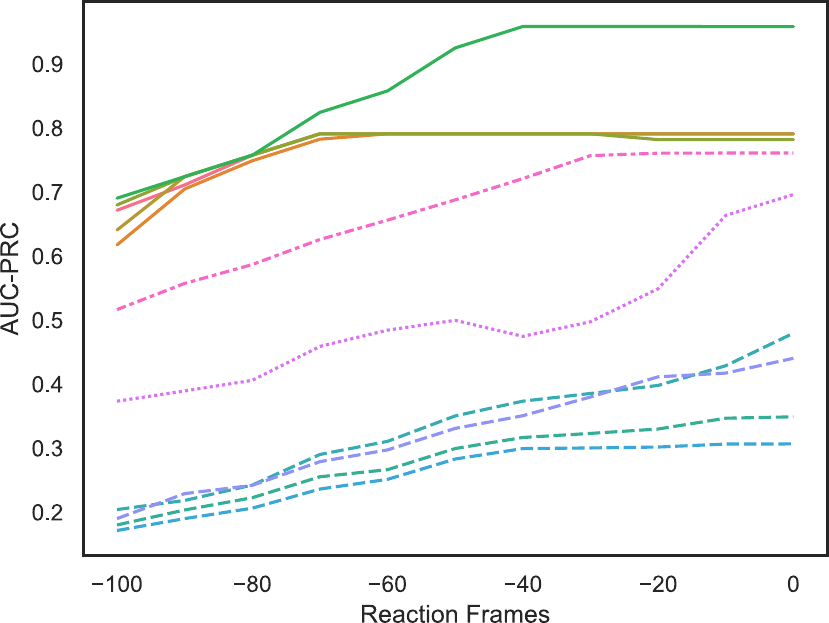}}
\vspace{-5pt}
\caption{AUC-PRC over reaction time for the LeoRover dataset}
\label{fig:ReactionAUCPRC}
\end{figure*}

Figure \ref{fig:ReactionAUCPRC} shows graphs with the different AUC-PRC scores (vertical axis) that would be obtained if we require the runtime monitors to raise an alarm a certain amount of frames \emph{before} the out-of-bounds happens (horizontal axis). For instance, the AUC-PRC at Reaction Frames $= -20$ is the measure that we would obtain in practice if we accounted for a delay of 20 frames from the moment an image is recorded until the oracle has processed it, the alarm is raised, and the healing action is taken. Since we process images at 10 FPS, Reaction Frames $= -20$ corresponds with roughly two seconds of latency. Note that the rightmost values (Reaction Frames $= 0$) correspond with the measures from Table \ref{table:ResultsLeoRoverAUC}. Note that the vertical axes are scaled to fit the data, and therefore the graphs have different scales.

Figures \ref{fig:ReactionAUCPRC:AnomalyCircuit-1} and \ref{fig:ReactionAUCPRC:AnomalyCircuit-2} show the AUC-PRC scores for the external filter-based anomalies in Circuit-1 and Circuit-2 respectively. We can observe that the performance of all approaches degrades very significantly if we account for 10 or 20 latency frames. In practice, this might mean that achieving results similar to the ones presented in the previous RQs may require very low latency communications and processing. Nevertheless, \tool and all the baselines seem to be affected by this in a very similar manner. The cause is likely to be the short reaction time between the moment the external anomaly happens and the out-of-bounds episode, as mentioned in the previous RQs.

On the other hand, Figures \ref{fig:ReactionAUCPRC:EnvironmentalAnomalyCircuit-1} and \ref{fig:ReactionAUCPRC:EnvironmentalAnomalyCircuit-2} show the AUC-PRC scores for the external environmental anomalies in Circuit-1 and Circuit-2 respectively. Similar to the external filter-based anomalies, we can see that the performance of all approaches degrades significantly after a few frames ,with \tool remaining the most effective for all reaction timeframes. We can see that the AUC-ROC of all the approaches is 0 at around 60 latency frames (or 6 seconds) in these datasets, since most of the environmental anomalies are very extreme and cause the DNN to drive out-of-bounds the fastest, meaning that the anomaly may not even have started 60 frames before the out-of-bounds.

Finally, Figures \ref{fig:ReactionAUCPRC:MutantCircuit-1} and \ref{fig:ReactionAUCPRC:MutantCircuit-2} show the AUC-PRC scores for the internal anomalies in Circuit-1 and Circuit-2 respectively. In this case, we can see that \tool's performance is almost completely unaffected even if we account for up to 40 frames (around four seconds) of latency. This should be realistically more than enough reaction time even if \tool's processing is performed in a separate computational unit communicating with the LeoRover. On the other hand, most baselines show a steady degradation in their effectiveness if any amount of latency is accounted for. Since SelfOracle derives its uncertainty scores purely from the DNN inputs (i.e., the camera images), it makes sense that most uncertain behaviors are identified shortly before the out-of-bounds occurs, which causes its efficiency to degrade if we account for some latency. Conversely, \tool does consider the DNN output, which can make it more sensitive to the DNN's internal behavior and allow it to identify mutants early. MC Dropout, which is a white-box approach, does show more effectiveness than the other baselines, particularly in Circuit-2, although not as much as \tool. Despite being a white-box approach, Ensemble does not seem to identify uncertain behaviors until shortly before the out-of-bounds occurs, as the graphs show a steady degradation of AUC-PRC comparable with SelfOracle's.

\subsubsection{Nvidia's Dave2 Dataset} \phantom \\

\begin{figure*}[ht]
\centering
\subfloat[External Anomalies]{\label{fig:ReactionAUCPRCDave2:Anomaly}\includegraphics[width=0.48\textwidth]{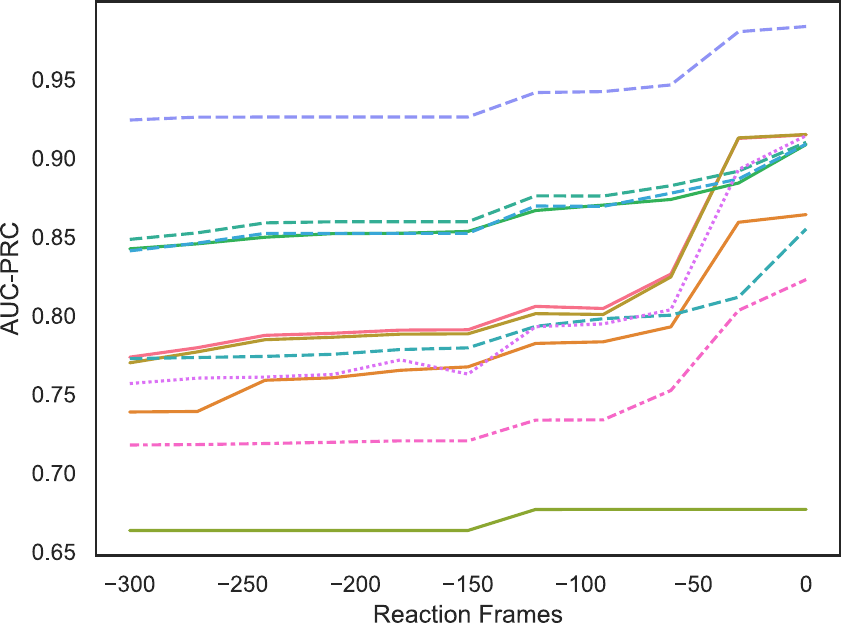}}
\,
\subfloat[Internal Anomalies]{\label{fig:ReactionAUCPRCDave2:Mutant}\includegraphics[width=0.48\textwidth]{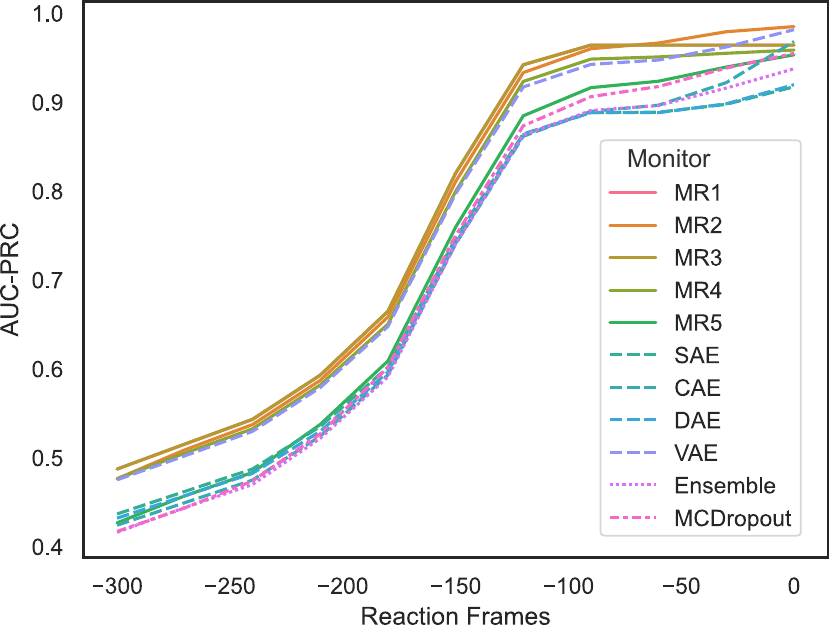}}
\vspace{-5pt}
\caption{AUC-PRC over reaction time for the Dave-2 dataset}
\label{fig:ReactionAUCPRCDave2}
\end{figure*}

Figure \ref{fig:ReactionAUCPRCDave2} shows graphs with the different AUC-PRC scores with different reaction timeframes for \tool and all the baselines. Unlike Figure \ref{fig:ReactionAUCPRC}, the horizontal axis goes from 0 to 300 reaction frames in increments of 30 frames, since the NVIDIA's Dave-2 case study datasets are recorded at approximately 30 frames per second instead of 10. This means that the graphs consider reaction times from 0 to 10 seconds before the out-of-bounds in this case study system as well.

Figure \ref{fig:ReactionAUCPRCDave2:Anomaly} shows the AUC-PRC results for the external anomaly datasets with different reaction periods. It is worth noting that in these datasets, unlike the LeoRover ones, the anomaly (fog, rain, snow or day/night) is active from the beginning of the simulation, which is why there are no steep slopes in this graph like those in the Anomaly datasets from Figure \ref{fig:ReactionAUCPRC}. Here, we can see that the effectiveness of MR1, MR2, MR3 and Ensemble degrades faster than the rest of the runtime monitors at around 60 frames (approximately 2 seconds) of reaction time. If more than 60 frames of reaction time is needed, we can see that SelfOracle's VAE is the most effective approach for any reaction timeframe in the graph, followed by SAE, DAE and MR5.

Figure \ref{fig:ReactionAUCPRCDave2:Mutant} shows the AUC-PRC results for the external anomaly datasets. Here, we can see that all the approaches have a very similar effectiveness with all the reaction timeframes considered in the graph, with \tool's MR2 and MR3 being the most effective.

\subsubsection*{Concluding Remark for RQ4} \phantom \\

Based on the evaluation of our approach and selected state-of-the-art techniques on two different case study systems, we can answer the fourth RQ as follows:

\begin{custombox}{RQ4}
In summary, we conclude that in the case of external anomalies in the LeoRover case study, the performance of \tool and the baselines degrades significantly if we account for a small latency. Conversely, in the case of internal anomalies in the LeoRover case study and all datasets in the Dave-2 case study, the performance of \tool and the baselines remains similar in most cases even with a latency of up to four seconds.
\end{custombox}

\subsection{RQ5 -- Frontier of Behaviors}

This RQ aims to evaluate the ability of \tool to avoid false positives under scenarios which are closer to the DNN's frontier of behaviors than the ideal conditions from the ``Nominal'' dataset. To this end, we collected datasets with anomalies that do not result in system misbehaviors for our LeoRover case study, which we describe in Section \ref{sec:Dataset:Passing}.

Table \ref{table:ResultsLeoRoverThresholdsPassing} shows the evaluation results for all the anomaly datasets with no misbehaviors, using the same thresholds as in Table \ref{table:ResultsLeoRoverThresholds}. Column ``External Filter Anomaly'' shows that, in Circuit-1, \tool's MR5 (Horizontal Flip) and MR1 (Reduce Brightness) obtain the highest number of FPs, with FPRs of 50\% and 23\% respectively, with MR4 (Blur), SelfOracle's CAE, and MR2 (Increase Contrast) also obtaining some false positives (FPR $<$10\%). In Circuit-2, however, FPRs are much lower, with only MR5 obtaining an FPR of 8\% and CAE obtaining an FPR of 6\%. On the other hand, Column “External Environ. Anomaly” shows that all of \tool's MRs have FPRs ranging from 11\% to 44\%, whereas SelfOracle's CAE has FPRs of 16\% and 20\%. Finally, Column “Internal Anomaly” shows that MR1 and MR4 have FPRs of up to around 20\%, while MR5 has an FPR of 53\% in Circuit-1. In comparison, CAE obtains an FPR of up to 16\%, and Ensemble obtains an FPR of up to 43\%.

These results indicate that \tool (specifically with MR1, MR4 and MR5) is the approach that is most prone to trigger alarms under various types of anomalies, even if they will not necessarily trigger misbehaviors of the system.

In order to measure the ability of the runtime monitors to identify misbehavior-inducing anomalies \emph{only}, we evaluate their ability to distinguish between the anomalies that cause misbehaviors and those that do not. Table \ref{table:ResultsLeoRoverAUCPassing} shows the threshold-independent AUC-PRC and AUC-ROC metrics, calculated with the non-misbehaving anomaly datasets and the corresponding misbehaving datasets. Column “External Filter Anomaly” shows that MR5 is the most effective at distinguishing between misbehaving and non-misbehaving anomalies in Circuit-2, whereas in Circuit-1, Ensemble achieves the highest AUC-PRC but MR3 achieves the highest AUC-ROC. On the other hand, Column “External Environ. Anomaly” shows that MR3 is the most effective in Circuit-1 in terms of both metrics, followed by Ensemble, whereas in Circuit-2, MR2 is the most effective, followed by MR1 and MR5. Finally, Column “Internal Anomaly” shows that MR5 is the most effective at distinguishing between misbehaving and non-misbehaving anomalies in all cases, and that all of the proposed MRs outperform all the baselines in this task.

\input{tables_results_RQ5}

\subsubsection*{Concluding Remark for RQ5} \phantom \\

Based on the evaluation of our approach and selected state-of-the-art techniques on the LeoRover case study system, the fifth RQ can be answered as follows:

\begin{custombox}{RQ5}
In summary, we conclude that, with the thresholds we have selected, \tool is the most sensitive approach to all types of anomalies, regardless of whether or not they can provoke misbehaviors. However, we also conclude that, given the right thresholds, \tool is the most effective at distinguishing between anomalies that will cause misbehaviors and those which will not.
\end{custombox}

\section{Discussion} \label{sec:discussion}

\subsection{Discussion and concluding remark}

Overall, we have found that \tool presents many benefits as a runtime monitor for DNNs over other approaches. On the one hand, most of the MRs implemented in \tool outperform all the baselines at identifying misbehaviors caused by internal anomalies in both of our case study systems. In both experiments, these internal anomalies were generated by applying DNN mutation operators to the driving DNNs, as proposed by Humbatova et al.~\cite{humbatova2021deepcrime}.

On the other hand, regarding external anomalies, we have implemented both image filter-based and environmental ones for our LeoRover case study, whereas in the Dave-2 case study they are implemented within the simulated environment. In these cases, we have obtained mixed results, as different approaches were more effective depending on the situation. In our LeoRover case study, we found that \tool is the most effective in most cases, but Ensemble could outperform it in one of the experiments. On the other hand, we found that SelfOracle's VAE is the most effective at identifying external anomalies in the Dave-2 case study.

It is apparent from the results that our method for calculating the uncertainty score thresholds for the runtime monitors results in FPs in many cases. For usages where FPs are highly undesirable, it might make sense to set a higher threshold by using a greater  margin from the maximum observed uncertainty in nominal conditions (e.g. 20\% instead of 10\%). On the other hand, it would also be possible to fine-tune the thresholds by performing further experimentation with different datasets for each specific runtime monitor. Finding an appropriate threshold is a complex problem that might require more research and specific solutions for different runtime monitors.

Regarding the assumptions that the different runtime monitors make, the MRs defined in \tool assume that the DNN has a degree of tolerance for minor image perturbations (MRs 1-4), and that the driving behavior is symmetric (MR5: Horizontal Flip). On the other hand, Ensemble assumes that the variance of the DNN outputs is small for nominal driving conditions, and MC Dropout assumes that the DNN model has a tolerance to deactivating a small percentage of the neurons. For our lane-keeping experiments, where some tolerance to input perturbations is expected from the models and there is a single specific correct output (with tolerance for small deviations), these approaches could in theory identify different kinds of anomalies affecting a lane-keeping DNN. SelfOracle is different than the rest, because it does not interact with the lane-keeping DNN, but instead tries to identify out-of-distribution inputs at runtime. This means that, while SelfOracle is expected to identify external anomalies, it should not be able to identify internal anomalies before they provoke a failure that can be observed in the input image (crash or out-of-bounds).

It is interesting that SelfOracle's CAE model achieves the best results among the four autoencoder models in our LeoRover case study, whereas in the evaluation from Stocco et al. it obtained the worst results \cite{stocco2020misbehaviour}. The relative ordering for the rest of the autoencoders from best to worst, is VAE, SAE, and lastly DAE, which does match the results obtaied by Stocco et al. \cite{stocco2020misbehaviour}. On the other hand, our results from our evaluation with the Dave-2 case study do show VAE achieving the best results. It is important to note that our LeoRover dataset comprises pictures from a physical environment, whereas SelfOracle's original evaluation employed images obtained from a simulator. CAE's image reconstruction process appears to be more sensitive to anomalies in real images than the rest of the autoencoders, but less so in images generated from a simulator.
Another possible factor might be the differences between the anomalies implemented in both case study systems: Upon visual inspection, the anomalies implemented in the Udacity simulation for the Dave-2 case study seem more subtle than both the external filter-based and external environmental anomalies implemented for our LeoRover case study. The anomalies in the Dave-2 case study are also present throughout the whole simulation, whereas in the LeoRover case study we introduce them abruptly.

Regarding resource consumption, we found that \tool is one of the cheapest approaches (in terms of CPU and RAM usage), since it only requires a basic image manipulation and an additional inference from the DNN. We hypothesize that, for more complex DNNs, \tool would become relatively cheaper than Ensemble and MC Dropout, since those approaches require more additional inferences from the DNN. On the other hand, the cost of SelfOracle should theoretically not change, since it does not interact with the DNN.

Regarding the MRs we implemented, in light of our evaluation results, it would make sense to employ a combination of MR5 (Horizontal Flip), MR2 (Increase Contrast) and MR3 (Add Noise) if possible, since these seem to cover most failures in our experiments. Nevertheless, it also seems that this selection can be narrowed down for each specific case study after some preliminary evaluation (only MR5 for LeoRover, and MR2 combined with MR3 for Dave-2).

In conclusion, \tool is capable of identifying various types of misbehavior-provoking anomalies that other approaches cannot at a reasonable cost, using the uncertainty scores computed from the proposed MRs. It is most effective at identifying internal anomalies, and can be complementary to other approaches that may be better suited for identifying external anomalies, such as SelfOracle.

\subsection{Threats to Validity}

We now discuss the main threats of our evaluation and how we tried to mitigate them.

\textbf{Internal validity:} The generated mutants and anomalies suppose potential internal validity threats in our study. For the mutants, we mitigated them by using some of the mutation operators recommended by Humbatova et al.~\cite{humbatova2021deepcrime}. As for the anomalies, we used the image corruptions and perturbations proposed by Hendrycks et al.~\cite{hendrycks2019robustness}, which are widely used for testing image-based DNNs. The employed DNN architecture and its training configuration could be another internal validity threat. In the case of the LeoRover, we mitigated it by employing the same architecture as suggested by the manufacturer. The other case study employs Nvidia's Dave-2 model, which has been widely used as object of study in prior related work \cite{pei2017deepxplore,tian2018deeptest,zhang2018deeproad,stocco2020misbehaviour,michelmore2020uncertainty,stocco2022thirdeye}. Moreover, we used the same architecture and parameters for the baseline of SelfOracle as its original paper~\cite{stocco2020misbehaviour}. The selected MRs 1 to 4 have some parameters (e.g., brightness level). Different parameters may lead to different results. To mitigate this threat, we verify that our parameters will result in valid follow-up inputs. We follow the definition provided by Riccio et al.~\cite{riccio2023and} to define a valid test input in the domain of DNN-based ADSs, i.e., the image is recognizable by a human expert.

\textbf{External validity:} The case study employed in the experimental evaluation could also pose a threat in terms of generalizability. We mitigated this threat by employing two different case studies, multiple circuits for the LeoRover case study, different anomaly types, and different DNN mutants, resulting in a diverse dataset. Moreover, the first employed case study system was a physical one, which reduces the threat posed by the gap between simulators and real ADSs. Meanwhile, the second case study system was based on a high-fidelity simulator based on a dataset provided in a prior paper~\cite{stocco2022thirdeye}, which helped us reduce the threat of potential input image distribution from actual road environments.

\textbf{Conclusion validity:} The employed baseline, i.e., SelfOracle~\cite{stocco2020misbehaviour}, employs neural networks, and as such, the training process is stochastic. To reduce the threat that the obtained results are by chance, we train each configuration of SelfOracle 10 times and provide average values in the evaluation.

%% file: tables_results.tex
\begin{table}[h]
\small
\setlength{\tabcolsep}{2pt}
\renewcommand{\arraystretch}{0.90}
\caption{LeoRover evaluation results for all oracles with Nominal and Anomalies \textbf{with misbehaviors} datasets with a 10\% threshold (average of 10 for SelfOracle), false positives and best results in boldface}
\resizebox{\textwidth}{!}{
\begin{tabular}{l rr| rrr rrr| rrr rrr| rrr rrr}
\toprule
 & \multicolumn{2}{c|}{Nominal} & \multicolumn{6}{c|}{External Filter Anomaly} & \multicolumn{6}{c|}{External Environ. Anomaly} & \multicolumn{6}{c}{Internal Anomaly} \\
 & \multicolumn{1}{c}{Circuit-1} & \multicolumn{1}{c|}{Circuit-2} & \multicolumn{3}{c}{Circuit-1} & \multicolumn{3}{c|}{Circuit-2} & \multicolumn{3}{c}{Circuit-1} & \multicolumn{3}{c|}{Circuit-2} & \multicolumn{3}{c}{Circuit-1} & \multicolumn{3}{c}{Circuit-2}\\
 & \multicolumn{1}{c}{FPR} & \multicolumn{1}{c|}{FPR} & \multicolumn{1}{c}{TPR} & \multicolumn{1}{c}{Prec.} & \multicolumn{1}{c}{F1} & \multicolumn{1}{c}{TPR} & \multicolumn{1}{c}{Prec.} & \multicolumn{1}{c|}{F1} & \multicolumn{1}{c}{TPR} & \multicolumn{1}{c}{Prec.} & \multicolumn{1}{c}{F1} & \multicolumn{1}{c}{TPR} & \multicolumn{1}{c}{Prec.} & \multicolumn{1}{c|}{F1} & \multicolumn{1}{c}{TPR} & \multicolumn{1}{c}{Prec.} & \multicolumn{1}{c}{F1} & \multicolumn{1}{c}{TPR} & \multicolumn{1}{c}{Prec.} & \multicolumn{1}{c}{F1}\\
\midrule
\textbf{\tool} &&&&&&&&&&&&&&&&&& \\
\hspace{5pt} MR1: Reduce Bright. & 0.00 & 0.00 & 0.15 & 1.00 & 0.27 & 0.33 & 1.00 & 0.49 & 0.33 & 1.00 & 0.50 & 0.33 & 1.00 & 0.50 & 0.83 & 1.00 & 0.91 & 0.83 & 1.00 & 0.91 \\
\hspace{5pt} MR2: Incr. Contrast & 0.00 & 0.00 & 0.06 & 1.00 & 0.11 & 0.04 & 1.00 & 0.07 & 0.22 & 1.00 & 0.36 & 0.33 & 1.00 & 0.50 & 0.77 & 1.00 & 0.87 & 0.77 & 1.00 & 0.87 \\
\hspace{5pt} MR3: Add Noise & 0.00 & 0.00 & 0.06 & 1.00 & 0.11 & 0.08 & 1.00 & 0.14 & 0.22 & 1.00 & 0.36 & 0.00 & 0.00 & 0.00 & 0.83 & 1.00 & 0.91 & 0.83 & 1.00 & 0.91 \\
\hspace{5pt} MR4: Blur & 0.00 & 0.00 & 0.04 & 1.00 & 0.07 & 0.12 & 1.00 & 0.21 & 0.00 & 0.00 & 0.00 & 0.22 & 1.00 & 0.36 & 0.83 & 1.00 & 0.91 & 0.83 & 1.00 & 0.91 \\
\hspace{5pt} MR5: Horizontal Flip & 0.00 & \textBF{0.10} & \textBF{0.35} & 1.00 & \textBF{0.51} & \textBF{0.65} & 0.92 & \textBF{0.76} & \textBF{0.44} & 1.00 & \textBF{0.62} & \textBF{0.56} & 0.62 & \textBF{0.59} & \textBF{1.00} & 1.00 & \textBF{1.00} & \textBF{1.00} & 0.91 & \textBF{0.96} \\
\textbf{SelfOracle} &&&&&&&&&&&&&&&&&& \\
\hspace{5pt} CAE & 0.00 & \textBF{0.00} & 0.17 & 1.00 & 0.29 & 0.27 & 0.99 & 0.41 & 0.18 & 1.00 & 0.30 & 0.11 & 0.60 & 0.19 & 0.12 & 0.70 & 0.20 & 0.09 & 0.79 & 0.16 \\
\hspace{5pt} DAE & 0.00 & 0.00 & 0.00 & 0.00 & 0.00 & 0.00 & 0.00 & 0.00 & 0.00 & 0.00 & 0.00 & 0.00 & 0.00 & 0.00 & 0.00 & 0.00 & 0.00 & 0.00 & 0.00 & 0.00 \\
\hspace{5pt} SAE & 0.00 & 0.00 & 0.05 & 0.50 & 0.09 & 0.06 & 0.50 & 0.10 & 0.00 & 0.00 & 0.00 & 0.06 & 0.50 & 0.10 & 0.00 & 0.00 & 0.00 & 0.00 & 0.00 & 0.00 \\
\hspace{5pt} VAE & 0.00 & 0.00 & 0.09 & 1.00 & 0.17 & 0.10 & 1.00 & 0.18 & 0.10 & 0.90 & 0.18 & 0.10 & 0.90 & 0.18 & 0.00 & 0.00 & 0.00 & 0.00 & 0.00 & 0.00 \\
\textbf{Ensemble} & \textBF{0.07} & \textBF{0.03} & 0.31 & 0.89 & 0.46 & 0.37 & 0.95 & 0.53 & 0.11 & 0.33 & 0.17 & 0.11 & 0.50 & 0.18 & 0.53 & 0.89 & 0.67 & 0.33 & 0.91 & 0.49 \\
\textbf{MC Dropout} & 0.00 & 0.00 & 0.00 & 0.00 & 0.00 & 0.00 & 0.00 & 0.00 & 0.00 & 0.00 & 0.00 & 0.00 & 0.00 & 0.00 & 0.50 & 1.00 & 0.67 & 0.33 & 1.00 & 0.50 \\
\bottomrule
\end{tabular}
} 
\label{table:ResultsLeoRoverThresholds}
\end{table}

\begin{table}[]
\small
\renewcommand{\arraystretch}{0.90}
\caption{LeoRover evaluation AUC results for all oracles and datasets \textbf{with misbehaviors} (average of 10 for SelfOracle), best results in boldface }
\resizebox{\textwidth}{!}{
\begin{tabular}{l rr rr| rr rr| rr rr}
\toprule
 & \multicolumn{4}{c|}{External Filter Anomaly} & \multicolumn{4}{c|}{External Environ. Anomaly} & \multicolumn{4}{c}{Internal Anomaly} \\
 & \multicolumn{2}{c}{Circuit-1} & \multicolumn{2}{c|}{Circuit-2} & \multicolumn{2}{c}{Circuit-1} & \multicolumn{2}{c|}{Circuit-2} & \multicolumn{2}{c}{Circuit-1} & \multicolumn{2}{c}{Circuit-2}\\
 \multicolumn{1}{c}{\textbf{AUC}} & \multicolumn{1}{c}{PRC} & \multicolumn{1}{c}{ROC} & \multicolumn{1}{c}{PRC} & \multicolumn{1}{c|}{ROC} & \multicolumn{1}{c}{PRC} & \multicolumn{1}{c}{ROC} & \multicolumn{1}{c}{PRC} & \multicolumn{1}{c|}{ROC} & \multicolumn{1}{c}{PRC} & \multicolumn{1}{c}{ROC} & \multicolumn{1}{c}{PRC} & \multicolumn{1}{c}{ROC}\\
\midrule
\textbf{\tool} &&&&&&&& \\
\hspace{5pt} MR1: Reduce Brightness & 0.688 & 0.439 & 0.831 & 0.727 & \textBF{0.486} & \textBF{0.746} & 0.572 & 0.767 & 0.791 & 0.833 & 0.791 & 0.833 \\
\hspace{5pt} MR2: Increase Contrast & 0.500 & 0.248 & 0.799 & 0.699 & 0.188 & 0.341 & \textBF{0.597} & \textBF{0.781} & 0.771 & 0.811 & 0.791 & 0.833 \\
\hspace{5pt} MR3: Add Noise & 0.578 & 0.346 & 0.870 & 0.821 & 0.339 & 0.535 & 0.390 & 0.780 & 0.791 & 0.833 & 0.791 & 0.833 \\
\hspace{5pt} MR4: Blur & 0.605 & 0.392 & 0.821 & 0.775 & 0.173 & 0.359 & 0.488 & 0.746 & 0.791 & 0.833 & 0.782 & 0.833 \\
\hspace{5pt} MR5: Horizontal Flip & \textBF{0.751} & \textBF{0.546} & 0.841 & 0.776 & 0.456 & 0.552 & 0.466 & 0.704 & \textBF{0.958} & \textBF{1.000} & \textBF{0.958} & \textBF{1.000} \\
\textbf{SelfOracle} &&&&&&&& \\
\hspace{5pt} CAE & 0.731 & 0.539 & 0.717 & 0.489 & 0.254 & 0.379 & 0.224 & 0.316 & 0.765 & 0.719 & 0.480 & 0.330 \\
\hspace{5pt} DAE & 0.517 & 0.214 & 0.568 & 0.322 & 0.155 & 0.259 & 0.208 & 0.327 & 0.376 & 0.269 & 0.307 & 0.146 \\
\hspace{5pt} SAE & 0.519 & 0.220 & 0.602 & 0.388 & 0.156 & 0.264 & 0.285 & 0.379 & 0.372 & 0.284 & 0.349 & 0.205 \\
\hspace{5pt} VAE & 0.532 & 0.256 & 0.685 & 0.522 & 0.170 & 0.315 & 0.179 & 0.281 & 0.512 & 0.393 & 0.440 & 0.311 \\
\textbf{Ensemble} & 0.695 & 0.462 & \textBF{0.882} & \textBF{0.848} & 0.125 & 0.174 & 0.301 & 0.600 & 0.692 & 0.715 & 0.696 & 0.748 \\
\textbf{MC Dropout} & 0.457 & 0.171 & 0.810 & 0.735 & 0.125 & 0.020 & 0.264 & 0.546 & 0.714 & 0.622 & 0.761 & 0.687 \\
\bottomrule
\end{tabular}
} 
\label{table:ResultsLeoRoverAUC}
\end{table}

%% file: tables_results_dave2.tex
\begin{table}[h]
\small
\setlength{\tabcolsep}{2pt}
\renewcommand{\arraystretch}{0.90}
\caption{Dave-2 evaluation results for all oracles and datasets with failures with a 10\% threshold (average of 10 for SelfOracle), false positives and best results in boldface}
\begin{tabular}{l r| rrr| rrr}
\toprule
 & \multicolumn{1}{c|}{Nominal}& \multicolumn{3}{c|}{External Anomaly} & \multicolumn{3}{c}{Internal Anomaly}\\
 & \multicolumn{1}{c|}{FPR} & \multicolumn{1}{c}{TPR} & \multicolumn{1}{c}{Prec.} & \multicolumn{1}{c|}{F1} & \multicolumn{1}{c}{TPR} & \multicolumn{1}{c}{Prec.} & \multicolumn{1}{c}{F1}\\
\midrule
\textbf{\tool} &&&&&&& \\
\hspace{5pt} MR1: Reduce Brightness & \textBF{0.10} & 0.34 & 0.96 & 0.50 & 0.64 & 0.99 & 0.78 \\
\hspace{5pt} MR2: Increase Contrast & \textBF{0.20} & 0.17 & 0.85 & 0.28 & \textBF{0.88} & 0.99 & \textBF{0.93} \\
\hspace{5pt} MR3: Add Noise & 0.00 & 0.34 & 1.00 & 0.51 & 0.64 & 1.00 & 0.78 \\
\hspace{5pt} MR4: Blur & \textBF{0.10} & 0.00 & 0.00 & 0.00 & 0.51 & 0.99 & 0.67 \\
\hspace{5pt} MR5: Horizontal Flip & \textBF{0.20} & 0.54 & 0.95 & 0.69 & 0.43 & 0.97 & 0.60 \\
\textbf{SelfOracle} &&&&&&& \\
\hspace{5pt} CAE & 0.00 & 0.18 & 1.00 & 0.29 & 0.48 & 1.00 & 0.63 \\
\hspace{5pt} DAE & 0.00 & 0.40 & 1.00 & 0.57 & 0.07 & 1.00 & 0.13 \\
\hspace{5pt} SAE & 0.00 & 0.43 & 1.00 & 0.60 & 0.07 & 1.00 & 0.13 \\
\hspace{5pt} VAE & \textBF{0.17} & \textBF{1.00} & 0.98 & \textBF{0.99} & 0.79 & 0.99 & 0.88 \\
\textbf{Ensemble} & 0.00 & 0.26 & 1.00 & 0.41 & 0.22 & 1.00 & 0.36 \\
\textbf{MC Dropout} & \textBF{0.20} & 0.17 & 0.85 & 0.28 & 0.46 & 0.98 & 0.63 \\
\bottomrule
\end{tabular}
\label{table:ResultsDave2Thresholds}
\end{table}

\begin{table}[]
\small
\renewcommand{\arraystretch}{0.90}
\caption{Dave-2 evaluation AUC results for all oracles and datasets with failures (average of 10 for SelfOracle), best results in boldface}
\resizebox{.7\textwidth}{!}{\begin{tabular}{l rr| rr}
\toprule
 & \multicolumn{2}{c|}{External Anomaly} & \multicolumn{2}{c}{Internal Anomaly} \\
  \multicolumn{1}{c}{\textbf{AUC}} & \multicolumn{1}{c}{PRC} & \multicolumn{1}{c|}{ROC} & \multicolumn{1}{c}{PRC} & \multicolumn{1}{c}{ROC} \\
\midrule
\textbf{\tool} &&&& \\
\hspace{5pt} MR1: Reduce Brightness & 0.915 & 0.742 & 0.964 & 0.531 \\
\hspace{5pt} MR2: Increase Contrast & 0.864 & 0.522 & \textBF{0.985} & \textBF{0.938} \\
\hspace{5pt} MR3: Add Noise & 0.915 & 0.740 & 0.964 & 0.531 \\
\hspace{5pt} MR4: Blur & 0.677 & 0.055 & 0.959 & 0.585 \\
\hspace{5pt} MR5: Horizontal Flip & 0.909 & 0.753 & 0.953 & 0.544 \\
\textbf{SelfOracle} &&&& \\
\hspace{5pt} CAE & 0.855 & 0.465 & 0.968 & 0.699 \\
\hspace{5pt} DAE & 0.909 & 0.649 & 0.920 & 0.294 \\
\hspace{5pt} SAE & 0.910 & 0.657 & 0.917 & 0.310 \\
\hspace{5pt} VAE & \textBF{0.984} & \textBF{0.900} & 0.982 & 0.884 \\
\textbf{Ensemble} & 0.914 & 0.732 & 0.938 & 0.419 \\
\textbf{MC Dropout} & 0.823 & 0.517 & 0.955 & 0.608 \\
\bottomrule
\end{tabular}}
\label{table:ResultsDave2AUC}
\end{table}

%% file: tables_results_RQ5.tex
\begin{table}[ht]
\centering
\setlength{\tabcolsep}{3pt}
\renewcommand{\arraystretch}{0.90}
\caption{LeoRover evaluation results for all oracles with Anomaly datasets without failures and a 10\% threshold (average of 10 for SelfOracle), false positives in boldface.}
\resizebox{\textwidth}{!}{
\begin{tabular}{l rr| rr| rr}
\toprule
 & \multicolumn{2}{c|}{External Filter Anomaly} & \multicolumn{2}{c|}{External Environ. Anomaly} & \multicolumn{2}{c}{Internal Anomaly} \\
 & \multicolumn{1}{c}{Circuit-1} & \multicolumn{1}{c|}{Circuit-2} & \multicolumn{1}{c}{Circuit-1} & \multicolumn{1}{c|}{Circuit-2} & \multicolumn{1}{c}{Circuit-1} & \multicolumn{1}{c}{Circuit-2} \\
 & \multicolumn{1}{c}{FPR} & \multicolumn{1}{c|}{FPR} & \multicolumn{1}{c}{FPR} & \multicolumn{1}{c|}{FPR} & \multicolumn{1}{c}{FPR} & \multicolumn{1}{c}{FPR} \\
\midrule
\textbf{\tool} &&&&&& \\
\hspace{5pt} MR1: Reduce Brightness & \textbf{0.23} & 0.00 & \textBF{0.22} & \textBF{0.33} & \textBF{0.23} & \textBF{0.23} \\
\hspace{5pt} MR2: Increase Contrast & \textBF{0.04} & 0.00 & 0.00 & \textBF{0.11} & \textBF{0.03} & 0.00 \\
\hspace{5pt} MR3: Add Noise & 0.00 & 0.00 & 0.00 & \textBF{0.11} & \textBF{0.07} & 0.00 \\
\hspace{5pt} MR4: Blur & \textBF{0.08} & 0.00 & \textBF{0.11} & \textBF{0.11} & \textBF{0.20} & \textBF{0.03} \\
\hspace{5pt} MR5: Horizontal Flip & \textBF{0.50} & \textBF{0.08} & \textBF{0.22} & \textBF{0.44} & \textBF{0.53} & \textBF{0.23} \\
\textbf{SelfOracle} &&&&&& \\
\hspace{5pt} CAE & \textBF{0.08} & \textBF{0.06} & \textBF{0.16} & \textBF{0.20} & \textBF{0.01} & \textBF{0.16} \\
\hspace{5pt} DAE & 0.00 & 0.00 & 0.00 & 0.00 & 0.00 & 0.00 \\
\hspace{5pt} SAE & 0.00 & 0.00 & 0.00 & 0.00 & 0.00 & 0.00 \\
\hspace{5pt} VAE & 0.00 & 0.00 & 0.00 & 0.00 & 0.00 & 0.00 \\
\textbf{Ensemble} & 0.00 & 0.00 & 0.00 & 0.00 & \textBF{0.43} & \textBF{0.13} \\
\textbf{MC Dropout} & 0.00 & 0.00 & 0.00 & 0.00 & 0.00 & 0.00 \\
\bottomrule
\end{tabular}
} 
\label{table:ResultsLeoRoverThresholdsPassing}
\end{table}

\begin{table}[]
\small
\renewcommand{\arraystretch}{0.90}
\caption{LeoRover evaluation AUC results for all oracles and datasets with failures versus without failures (average of 10 for SelfOracle), best results in boldface}
\resizebox{\textwidth}{!}{\begin{tabular}{l rr rr| rr rr| rr rr}
\toprule
 & \multicolumn{4}{c|}{External Filter Anomaly} & \multicolumn{4}{c|}{External Environ. Anomaly} & \multicolumn{4}{c}{Internal Anomaly} \\
 & \multicolumn{2}{c}{Circuit-1} & \multicolumn{2}{c|}{Circuit-2} & \multicolumn{2}{c}{Circuit-1} & \multicolumn{2}{c|}{Circuit-2} & \multicolumn{2}{c}{Circuit-1} & \multicolumn{2}{c}{Circuit-2}\\
 \multicolumn{1}{c}{\textbf{AUC}} & \multicolumn{1}{c}{PRC} & \multicolumn{1}{c}{ROC} & \multicolumn{1}{c}{PRC} & \multicolumn{1}{c|}{ROC} & \multicolumn{1}{c}{PRC} & \multicolumn{1}{c}{ROC} & \multicolumn{1}{c}{PRC} & \multicolumn{1}{c|}{ROC} & \multicolumn{1}{c}{PRC} & \multicolumn{1}{c}{ROC} & \multicolumn{1}{c}{PRC} & \multicolumn{1}{c}{ROC}\\
\midrule
\textbf{\tool} &&&&&&&& \\
\hspace{5pt} MR1: Reduce Brightness & 0.578 & 0.373 & 0.880 & 0.799 & 0.538 & 0.642 & 0.518 & 0.562 & 0.791 & 0.833 & 0.791 & 0.833 \\
\hspace{5pt} MR2: Increase Contrast & 0.584 & 0.356 & 0.828 & 0.754 & 0.355 & 0.469 & \textBF{0.562} & \textBF{0.611} & 0.779 & 0.822 & 0.791 & 0.833 \\
\hspace{5pt} MR3: Add Noise & 0.756 & \textBF{0.599} & 0.843 & 0.782 & \textBF{0.639} & \textBF{0.846} & 0.350 & 0.481 & 0.791 & 0.833 & 0.791 & 0.833 \\
\hspace{5pt} MR4: Blur & 0.578 & 0.395 & 0.840 & 0.777 & 0.420 & 0.543 & 0.391 & 0.531 & 0.791 & 0.833 & 0.782 & 0.833 \\
\hspace{5pt} MR5: Horizontal Flip & 0.612 & 0.396 & \textBF{0.896} & \textBF{0.864} & 0.527 & 0.537 & 0.511 & 0.549 & \textBF{0.958} & \textBF{1.000} & \textBF{0.958} & \textBF{1.000} \\
\textbf{SelfOracle} &&&&&&&& \\
\hspace{5pt} CAE & 0.738 & 0.550 & 0.814 & 0.726 & 0.328 & 0.393 & 0.398 & 0.347 & 0.598 & 0.531 & 0.393 & 0.278 \\
\hspace{5pt} DAE & 0.645 & 0.464 & 0.673 & 0.544 & 0.396 & 0.321 & 0.373 & 0.511 & 0.522 & 0.402 & 0.390 & 0.326 \\
\hspace{5pt} SAE & 0.641 & 0.456 & 0.714 & 0.597 & 0.381 & 0.370 & 0.384 & 0.528 & 0.533 & 0.414 & 0.426 & 0.344 \\
\hspace{5pt} VAE & 0.589 & 0.365 & 0.755 & 0.641 & 0.359 & 0.417 & 0.383 & 0.410 & 0.526 & 0.436 & 0.471 & 0.385 \\
\textbf{Ensemble} & \textBF{0.774} & 0.580 & 0.887 & 0.840 & 0.584 & 0.691 & 0.329 & 0.370 & 0.514 & 0.605 & 0.523 & 0.659 \\
\textbf{MC Dropout} & 0.600 & 0.389 & 0.883 & 0.807 & 0.363 & 0.469 & 0.271 & 0.222 & 0.715 & 0.618 & 0.746 & 0.663 \\
\bottomrule
\end{tabular}}
\label{table:ResultsLeoRoverAUCPassing}
\end{table}

%% file: RelatedWork.tex
Automated test generation techniques have been the most widely investigated technique to identify unexpected driving scenarios of ADSs. This has been targeted from different perspectives, including techniques to maximize neuron coverage~\cite{pei2017deepxplore}, surrogate-assisted search-based test generation~\cite{abdessalem2018testing,abdessalem2018testing2,haq2022efficient,mullins2018adaptive}, techniques based on reinforcement learning \cite{haq2023many,lu2022learning} and procedural content generation to generate virtual roads~\cite{gambi2019automatically}. These studies focus on detecting unexpected scenarios at design-time, before deploying the system in opration. Conversely, our approach focuses on proposing a run-time monitoring technique to build a safety envelope over a DNN to assess its level of dependability in operation. 

In this context, i.e., run-time monitoring of DNN-enabled ADSs, black-box~\cite{stocco2020misbehaviour,stocco2020misbehaviour} and white-box~\cite{stocco2022thirdeye} approaches have been recently proposed, as discussed throughout the paper. Our approach falls in the scope of the black-box ones. In such an scope, frameworks such as SelfOracle~\cite{stocco2020misbehaviour} and DeepGuard~\cite{hussain2022deepguard} train auto-encoders with the same images used for training the ADS, and use the reconstruction error of these auto-encoders as an uncertainty score. Unlike existing black-box approaches, \tool does not require any training, as it relies on metamorphic relations to generate follow-up inputs and subsequently obtain a confidence score through output relations. Moreover, \tool overcomes the limitation identified by Stocco et al.~\cite{stocco2022thirdeye} for black-box techniques, i.e., failing to detect faults caused by an inadequate training or by bugs at the DNN model level, as we demonstrated in our evaluation.

Our approach is largely inspired by MT, a technique that has been found effective at alleviating the test oracle problem~\cite{segura2016survey}. This technique has also been explored in the context of ADSs~\cite{segura2016survey,tian2018deeptest,zhou2019metamorphic,zhang2018deeproad,deng2021bmt,deng2022declarative}. Tian et al.~\cite{tian2018deeptest} proposed DeepTest, a testing framework to identify erroneous behaviors of DNN-based self-driving cars through MR-based test oracles. Zhang et al.~\cite{zhang2018deeproad} proposed DeepRoad, an ADS testing framework that leverages Generative Adversarial Networks (GANs) to automatically generate realistic follow-up images simulating various weather conditions, such as snow or rain. Deng et al.~\cite{deng2021bmt} propose a Behaviour-Driven Development (BDD) workflow in which MRs are expressed in a declarative manner by the test engineers. The framework converts declared MRs to executable oracles in addition to the synthesis of test inputs for each MR~\cite{deng2021bmt}. The same team later proposed a rule-based declarative framework that leverages natural language processing (NLP) to extract MRs from predicates written in natural language~\cite{deng2022declarative}. Unlike all these studies, which focus on design-time testing of ADSs, our approach leverages MRs for operational purposes with the goal of detecting potential misbehaviors and take corresponding healing actions in run-time.

To the best of our knowledge, there are only three papers that have proposed the use of MT at run-time~\cite{murphy2009automatic,murphy2009metamorphic,bell2015metamorphic}. These papers introduce the concept of \emph{Metamorphic Runtime Checking}, which is the analogous of runtime assertion checking for MRs \cite{bell2015metamorphic}. Similar to our work, their approach also involves intercepting inputs at runtime and executing follow-up test inputs to check MRs. In contrast to our work, which aims to monitor ADSs, they apply this technique to monitor generic software applications \cite{murphy2009automatic} and their individual functions \cite{murphy2009metamorphic,bell2015metamorphic}. Their approach involves creating a ``sandbox'' in order to execute the follow-up test cases, which typically involves forking the process, in order to ensure identical state from the original execution \cite{bell2015metamorphic}. This sandboxing process is unnecessary for \tool, since DNNs are generally isolated components which do not depend on external state. Given the application domain, the monitors they generate employ assertions (pass or fail). Unlike these approaches, we levarage metamorphic relations with the goal of measuring the uncertainty of the ADS controller in an efficient and effective way. \tool later employs these quantitative uncertainty scores in an auto-regressive filter to consider past values and raise a flag in case this value exceeds a predefined threshold. To the best of our knowledge, this paper is the first that aims at adapting MT to run-time monitoring in a context of AVs and CPSs.

%% file: Conclusion.tex
We propose \tool, a novel run-time monitoring approach that leverages ideas from Metamorphic Testing to estimate the confidence of ADS controllers. Our evaluation, which uses a small-scale physical ADS and a different ADS in a simulated environment, shows that \tool is capable of effectively anticipating failures caused by either internal or external anomalies before the vehicle goes out-of-bounds. Indeed, our approach outperformed or at least showed a performance comparable to the employed baselines (i.e., SelfOracle~\cite{stocco2020misbehaviour} Ensemble-based techniques and MC Dropout).

Future work includes comparing our approach with other uncertainty or confidence estimators, such as those based on eXplainable AI (XAI) techniques. Furthermore, we also plan to evaluate the impact of combining these techniques with different self-healing mechanisms, such as slowing down the vehicle when facing a sudden peak of the uncertainty score. Finally, we consider that specific methods to establish appropriate thresholds for each MR should be studied further.

\section*{Replication Packages: }We make all our artifacts available as follows:

\begin{itemize}
    \item Our dataset can be found at: 
    \url{https://doi.org/10.5281/zenodo.10716933}
    \item Our code is available at: \url{https://github.com/jonayerdi/marmot}
\end{itemize}

%% file: appendix1.tex
\begin{figure*}[h]
    \centering
    \subfloat[MarMot MR1: Reduce Brightness]{%
        \includegraphics[width=0.3\textwidth]{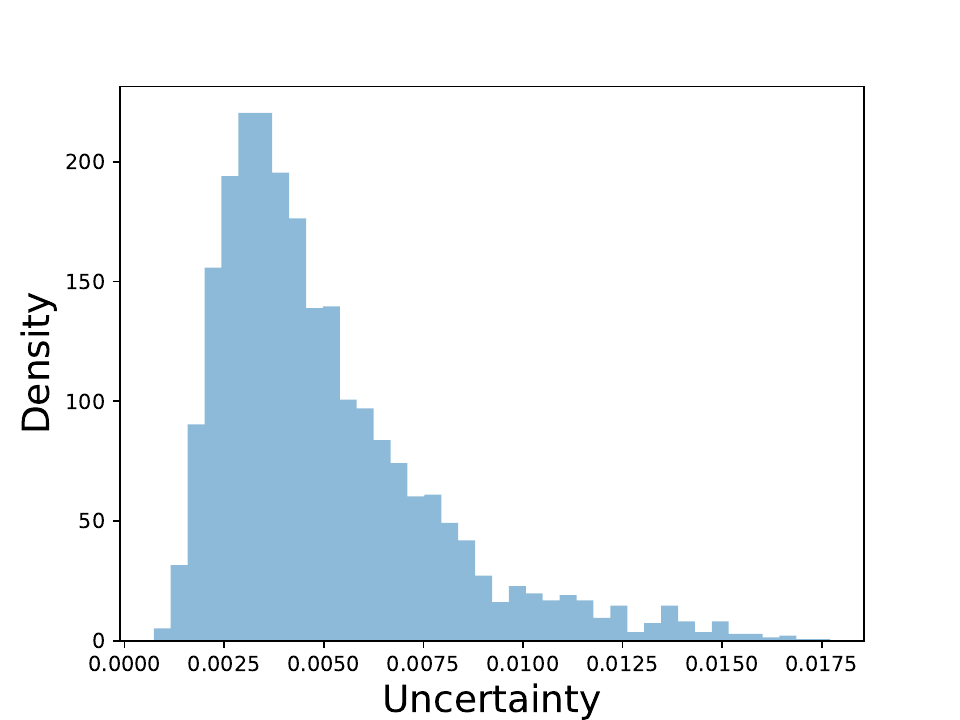}}\,
    \subfloat[MarMot MR2: Increase Contrast]{%
        \includegraphics[width=0.3\textwidth]{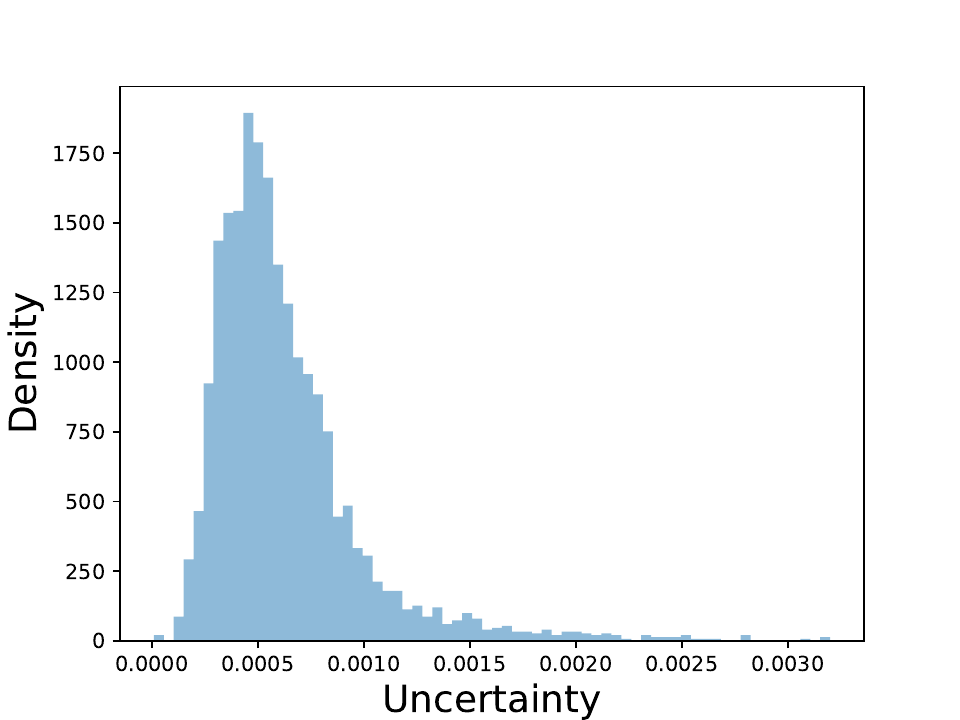}}\,
    \subfloat[MarMot MR3: Add Noise]{%
        \includegraphics[width=0.3\textwidth]{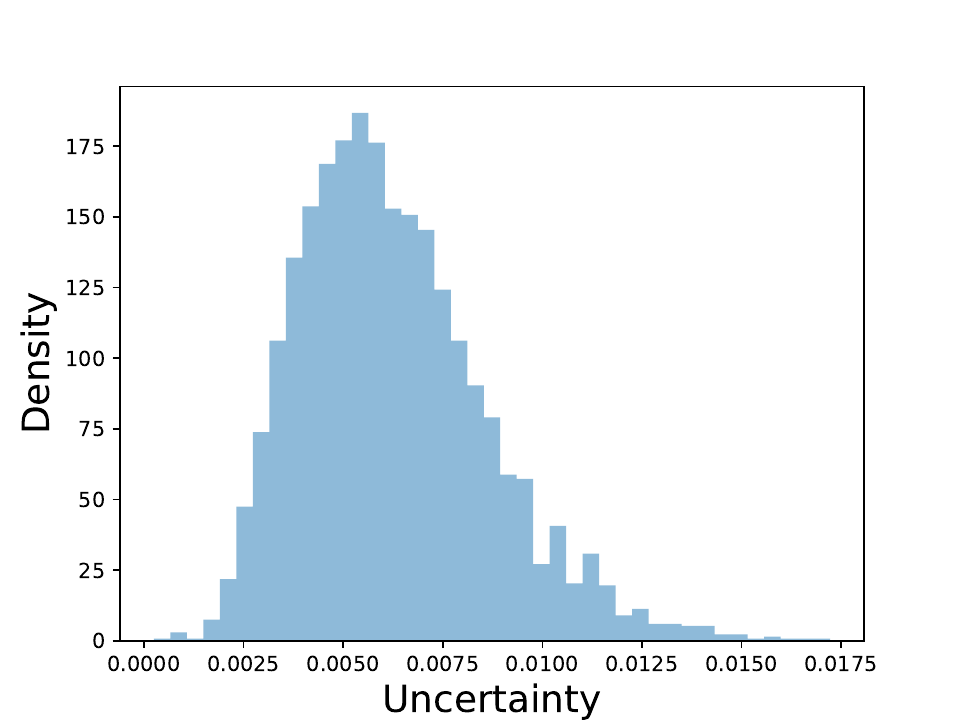}}\,
    \subfloat[MarMot MR4: Blur]{%
        \includegraphics[width=0.3\textwidth]{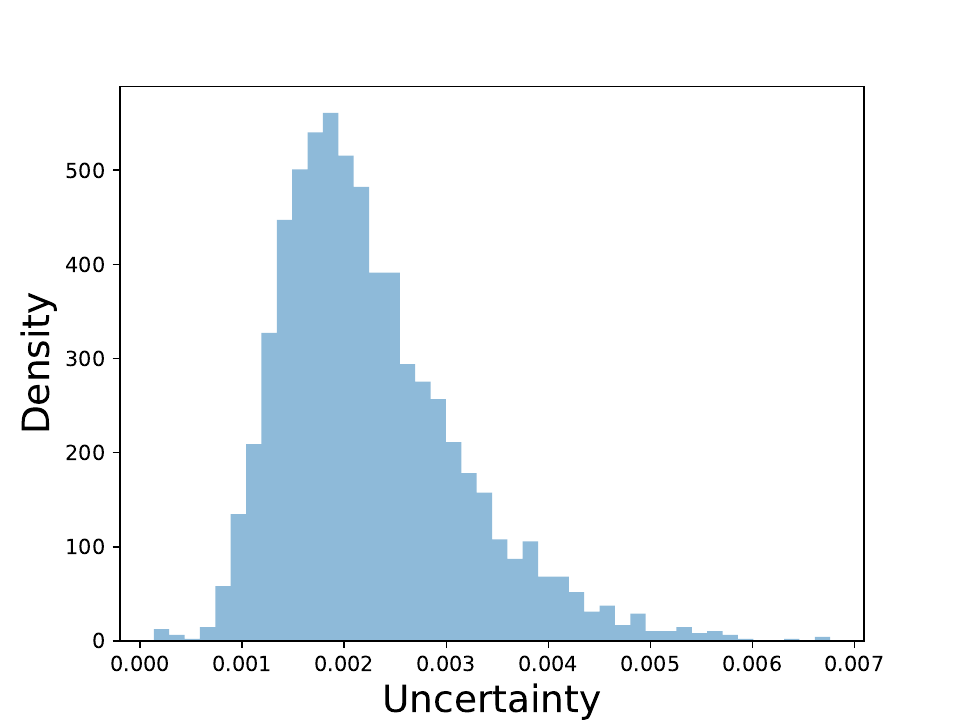}}\,
    \subfloat[MarMot MR5: Horizontal Flip]{%
        \includegraphics[width=0.3\textwidth]{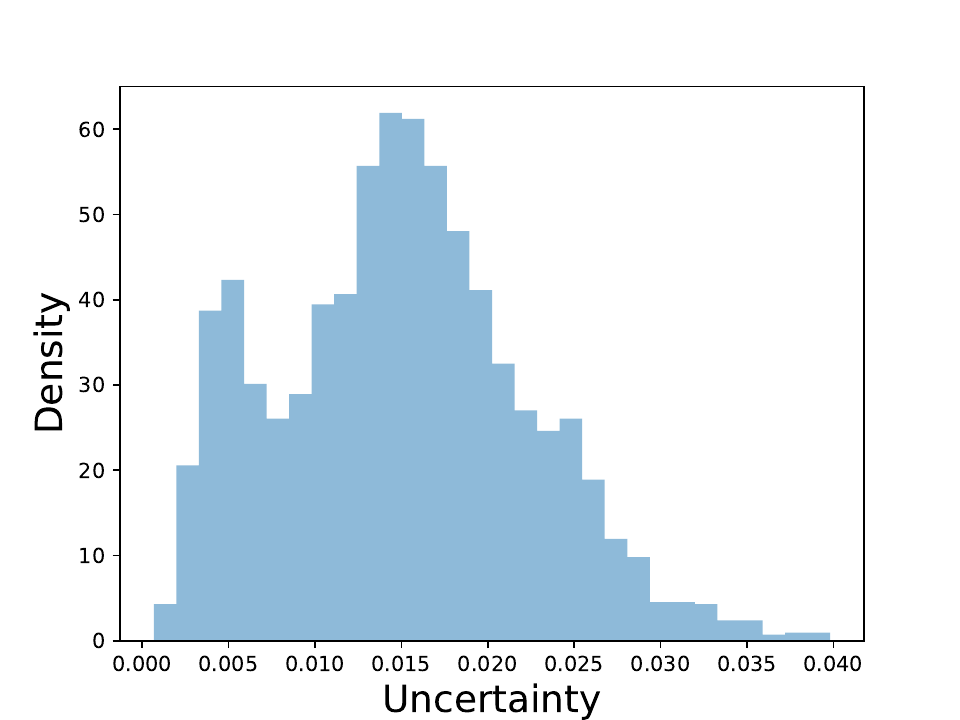}}\,
    \subfloat[Ensemble]{%
        \includegraphics[width=0.3\textwidth]{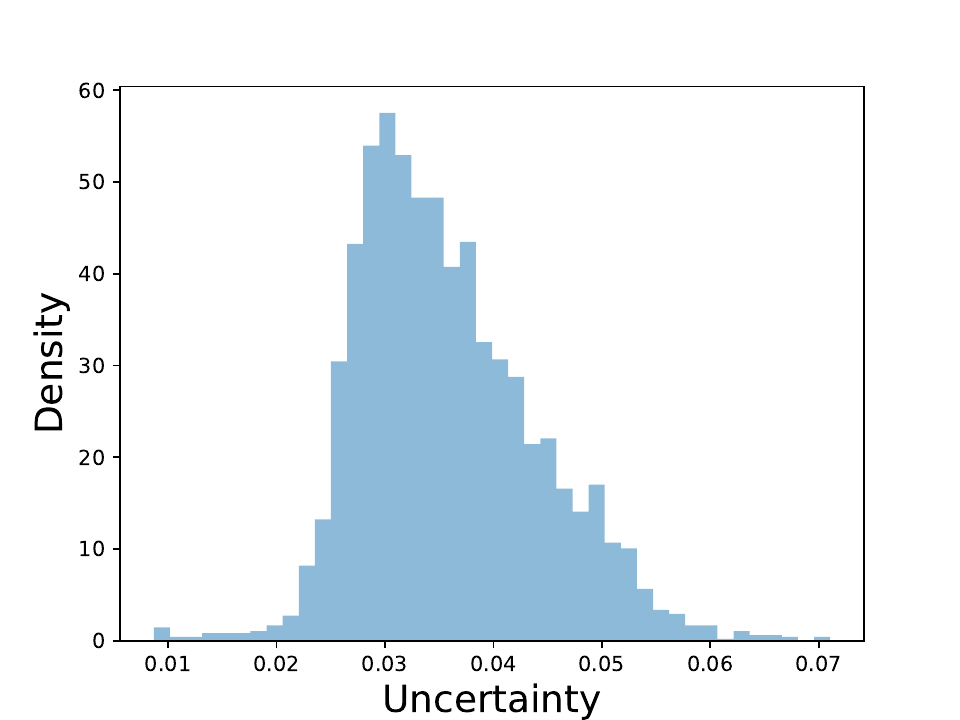}}\,
    \subfloat[SelfOracle CAE]{%
        \includegraphics[width=0.3\textwidth]{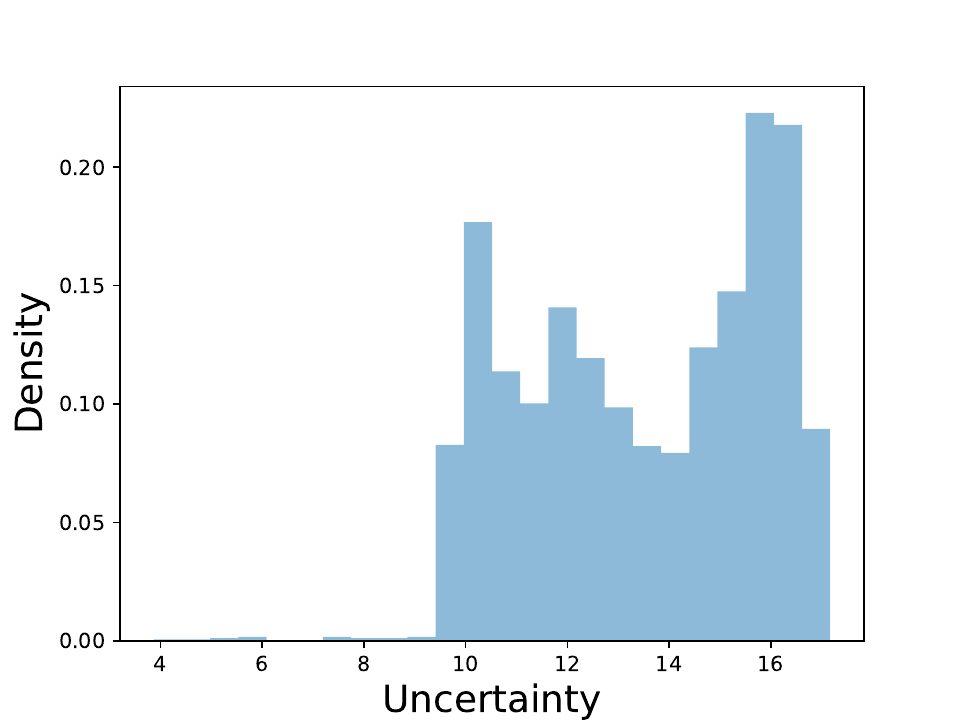}}\,
    \subfloat[SelfOracle DAE]{%
        \includegraphics[width=0.3\textwidth]{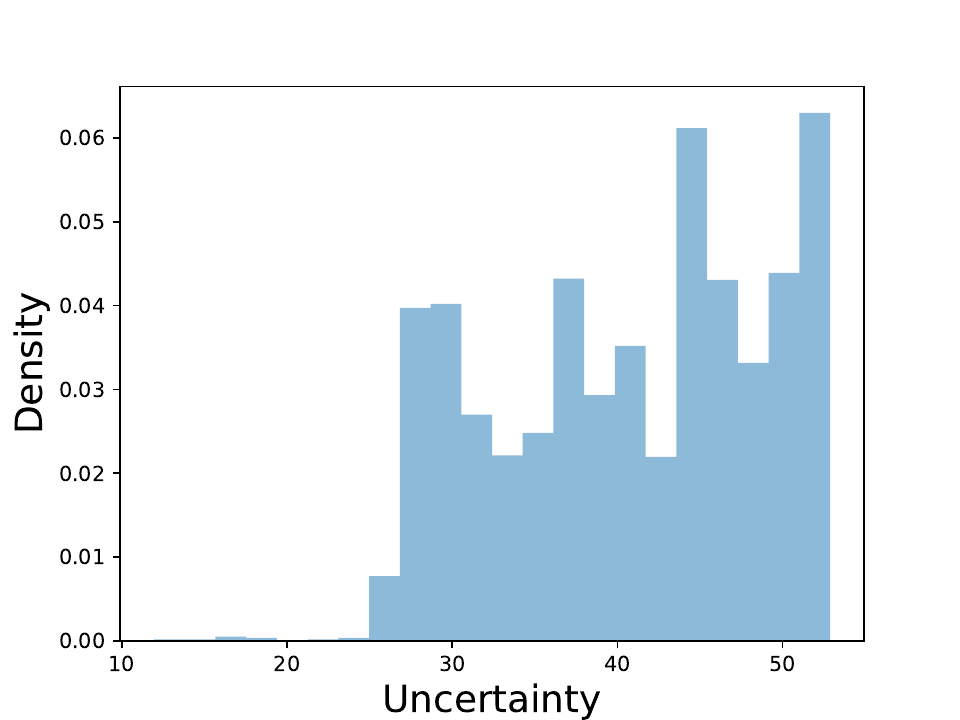}}\,
    \subfloat[SelfOracle SAE]{%
        \includegraphics[width=0.3\textwidth]{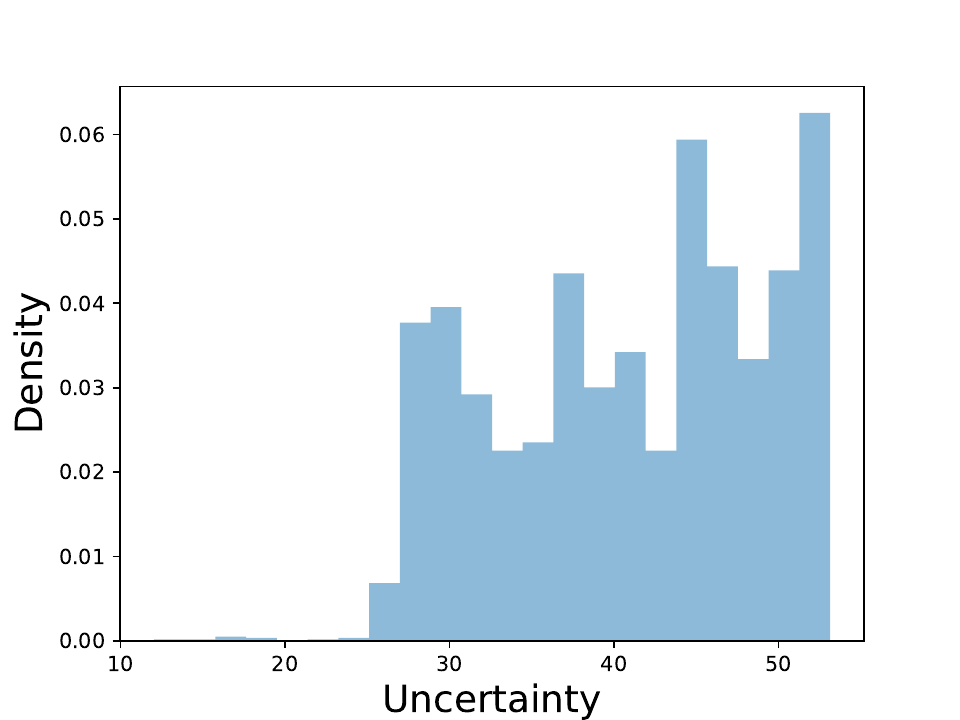}}\,
    \subfloat[SelfOracle VAE]{%
        \includegraphics[width=0.3\textwidth]{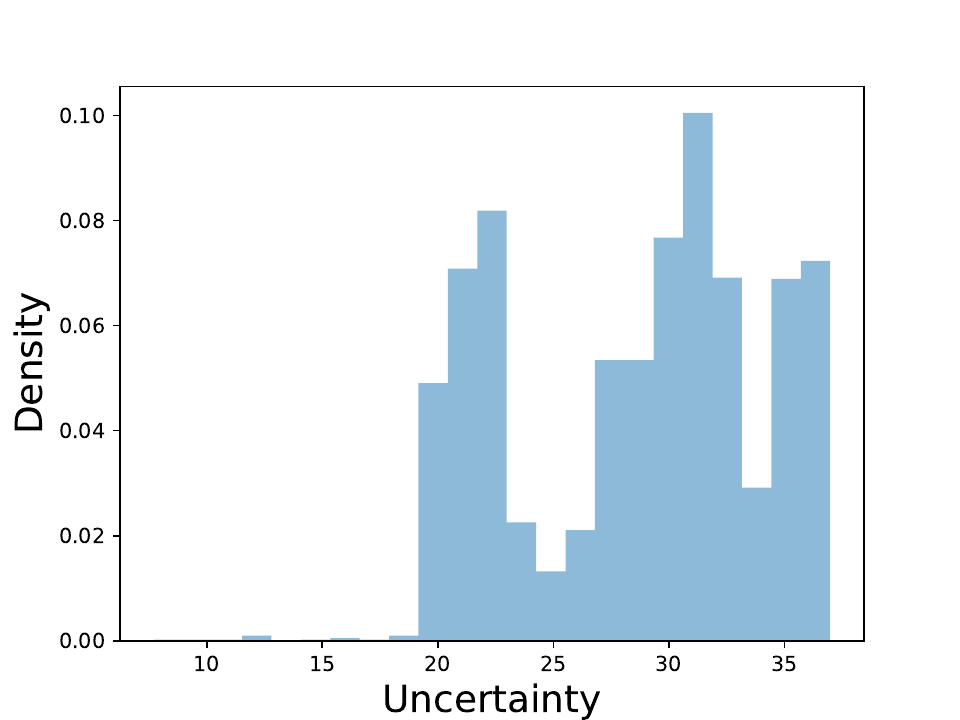}}\,
    \caption{Distribution of uncertainty scores under nominal conditions}
    \label{fig:threshold}
\end{figure*}

As discussed in Section~\ref{sec:Evaluation:Configuration:Thresholds}, the studied approaches require an uncertainty threshold value to determine whether the system is running under normal conditions. In order to set this threshold for each of the approaches, we observed 7 recordings of nominal driving conditions from Circuit-1. Figure~\ref{fig:threshold} shows the distribution of uncertainty scores per approach during these recordings.

